\documentclass[namedreferences]{solarphysics}

\usepackage[optionalrh]{spr-sola-addons} 

\usepackage{graphicx}        
\usepackage{color}           
\usepackage{soul}
\usepackage{booktabs}
\usepackage{amsmath}
\DeclareMathAlphabet\mathbfcal{OMS}{cmsy}{b}{n}

\renewcommand{\vec}[1]{{\mathbfit #1}}

\chardef\us=`\_

\usepackage[utf8]{inputenc}

\renewcommand{\vec}[1]{{\mathbf #1}}

\newcommand{\derivp}[2]{\frac{{\mathrm \partial} #1}{{\mathrm \partial} #2}}

\newcommand{\B}{{\textbf{B}}}

\begin{document}

\begin{article}

\begin{opening}

\title{Study of Reconnection Dynamics and Plasma Relaxation in MHD simulation of a Solar Flare}

\author[addressref={aff1,aff2},email={satyam@prl.res.in}]{\inits{S.}\fnm{Satyam}~\lnm{Agarwal}}
\author[addressref=aff1,email={ramit@prl.res.in}]{\inits{R.}\fnm{Ramit}~\lnm{Bhattacharyya}}
\author[addressref={aff3,aff4,aff5},email={yangshb@nao.cas.cn}]{\inits{S.}\fnm{Shangbin}~\lnm{Yang}}

\address[id=aff1]{Udaipur Solar Observatory, Physical Research Laboratory, Udaipur 313 001, India}
\address[id=aff2]{Indian Institute of Technology Gandhinagar, Palaj, Gujarat, India}
\address[id=aff3]{Key Laboratory of Solar Activity, National Astronomical Observatories, Chinese Academy of Sciences, 100101 Beijing, People's Republic of China}
\address[id=aff4]{School of Astronomy and Space Sciences, University of Chinese Academy of Sciences, 100049 Beijing, People's Republic of China}
\address[id=aff5]{Key Laboratory of Solar Activity and Space Weather, 100190 Beijing, People’s Republic of China}
\runningauthor{S. Agarwal et al.}
\runningtitle{Study of Reconnection Dynamics and Plasma Relaxation in MHD simulation of a Solar Flare}

\begin{abstract}

Self-organization in continuous systems is associated with dissipative processes. In particular, for magnetized plasmas, it is known as magnetic relaxation, where the magnetic energy is converted into heat and kinetic energy of flow through the process of magnetic reconnection. An example of such a system is the solar corona, where reconnection manifests as solar transients like flares and jets. Consequently, toward investigation of plasma relaxation in solar transients, we utilize a novel approach of data-constrained MHD simulation for an observed solar flare. The selected active region NOAA 12253 hosts a GOES M1.3 class flare. The investigation of extrapolated coronal magnetic field in conjunction with the spatiotemporal evolution of the flare reveals a hyperbolic flux tube (HFT), overlying the observed brightenings. MHD simulation is carried out with the EULAG-MHD numerical model to explore the corresponding reconnection dynamics. The overall simulation shows signatures of relaxation. For a detailed analysis, we consider three distinct sub-volumes. We analyze the magnetic field line dynamics along with time evolution of physically relevant quantities like magnetic energy, current density, twist, and gradients in magnetic field. In the terminal state, none of the sub-volumes are seen to reach a force-free state, thus remaining in non-equilibrium, suggesting the possibility of further relaxation. We conclude that the extent of relaxation depends on the efficacy and duration of reconnection, and hence, on the energetics and time span of the flare. 


\end{abstract}


\end{opening}

%


\section{Introduction}
\label{sec1}

Continuous dissipative systems exhibit self-organization (\citealp{1985H}) by evolving preferentially toward states characterized by long-range ordering in one physical variable and short-range disorder in other variables. In the context of magnetized plasmas, self-organization is generally termed as plasma relaxation wherein the magnetofluid relaxes toward a minimum energy state while preserving appropriate physical variables. This process is quantitatively understood by forward cascade in one variable---which gets dissipated at the smallest scales in the system while, inverse cascade of other variables generates the long range order. A qualitative analysis of the phenomenon is also possible by relying on the principle of selective decay (\citealp{1980M}). The principle effectively compares the decay rates of two or more variables in the presence of dissipation and the relaxed state is obtained by a constrained minimization of the fastest decaying variable while treating the slower decaying variables as invariants (\citealp{Orto}).

The constrained minimization can be explored by briefly revisiting the \citet{1958W} theory. There the relaxed state is obtained by minimizing the net magnetic energy of a magnetic flux tube while keeping its volume integrated magnetic helicity: a measure of magnetic topology, invariant. The relaxed state is characterized by a volume current density ($\textbf{J}$) to be entirely parallel to the magnetic field ($\textbf{B}$) such that the Lorentz force is zero ($\textbf{J}\times\textbf{B}=0$)---earning the moniker ``Force-Free State". Incidentally, the force-free state also describes a magnetostatic equilibrium for a low $\beta$ plasma, where the magnetic tension force is balanced by the magnetic pressure force. The proportionality factor between $\textbf{J}$ and $\textbf{B}$, although constant for a given flux tube, can vary over different flux tubes and hence, is a function of position. The equations representing the Woltjer state are

\begin{align}
&\nabla\times\textbf{B}=\alpha(\textbf{r})\textbf{B}, \label{ff1}\\
&\nabla \alpha(\textbf{r})\cdot\textbf{B}=0, \label{ff2}
\end{align}

\noindent where $\alpha(\textbf{r})$ represents the magnetic circulation per unit flux (\citealp{2012PARKER}). The second equation is necessary to impose the solenoidality of $\textbf{B}$. Together, these two equations are called nonlinear force-free equations and are used to describe various physical systems including the solar corona. 

J. B. Taylor further conjectured that in an isolated and slightly resistive plasma, magnetic reconnection between different flux tubes will homogenize $\alpha(\textbf{r})$ and plasma pressure (if any), resulting in a relaxed state obtained by a minimization of the global magnetic energy

\begin{equation}
    \label{magen}
    W=\displaystyle\int_V \frac{|\B|^2}{2}\,\mathrm{d}^{3}x,
\end{equation}
while keeping the global magnetic helicity
\label{hel}
\begin{equation}
    \label{held}
    H=\displaystyle\int_{V}\mathbf{A}\cdot\mathbf{B}\,\mathrm{d}^{3}x
\end{equation}

\noindent invariant (\citealp{1974T}, \citeyear{1986T}, \citeyear{2000T}). Notably, $\mathbf{A}$ is the vector potential and the volume $V$ encompasses the whole system domain. The relaxed state is described by the linear force-free field (\citealp{1957SC}), satisfying

\begin{equation}
\label{lfff}
\nabla\times\mathbf{B}=\alpha\mathbf{B},
\end{equation}

\noindent with a constant $\alpha$. Notably, relative to the global magnetic energy, the treatment of global magnetic helicity as an invariant is consistent with the selective decay principle---as demonstrated in \citet{1984Berger}, \citet{1988Br}, \citet{1998AB}, and nicely summarized in \citet{Ybook2020}. Since the conservation of global magnetic helicity is central to Taylor relaxation, it merits further discussion. Notably, the helicity quantifies the interlinking and knotting between different magnetic field lines along with twisting and kinking of a given filed line \citep{1984B}. Importantly, magnetic helicity is a pseudo-scalar because of its gauge dependency. For example, the definition in equation (\ref{held}) is gauge independent only for an isolated system with no magnetic field line cutting across its boundaries. For open systems, the definition of magnetic helicity involves reference fields along with apt boundary condition to make it gauge invariant---see \citet{Ybook2020} for details. 

Both the Woltjer and Taylor relaxed states focus only on magnetic properties of the plasma and do not include other variables like the plasma flow, kinetic pressure, or dissipation rates. Inclusions of these variables are possible within the framework of 
two-fluid magnetohydrodynamics (MHD). Generally in the two fluid formalism, the minimizer incorporates plasma flow whereas the invariants are either the generalized ion and electron helicities \citep{1997SI} or their derivatives \citep{Ramit2004}, and can even include total (magnetic +kinetic) energy \citep{IM2002}. The obtained relaxed states are always ``flow-coupled", i.e the plasma flow and magnetic field are interrelated. To avoid any confusion between the Woltjer or Taylor relaxed states with the two-fluid relaxed states, hereafter the former is called magnetic relaxation which, also emphasizes on its magnetic nature. Relevantly, \citet{2010Y} proposed 
the existence of an additional constraint along with magnetic helicity, namely the topological degree of field line mapping (also see 
\citealp{2015Y}, \citealp{2021Y}) to obtain relaxed states. The relaxed states turned out to be either linear force-free or nonlinear force-free depending on the topological degree.

One of the essential characteristics of magnetic relaxation is the dissipation of magnetic energy through reconnection. Relevantly, in solar coronal transients such as flares, jets, and coronal mass ejections, a fraction of the stored magnetic energy is dissipated in the form of heat and kinetic energy of charged particles through the process of magnetic reconnection (\citealp{2011S}, \citealp{2016Z}, \citealp{2021Li}). The reconnection assisted decay of magnetic energy along with the wealth of multiwavelength observations of these transients from space-based observatories, make them a suitable testbed to explore magnetic relaxation in nature. In this context, \citet{2003N} analyzed several flare-productive active regions and found their time evolution to be tending towards a linear force-free state. A similar result was found by \citet{2013M}, who investigated the pre-flare and post-flare coronal magnetic fields in active region NOAA 10953. The authors determined the post-flare configuration as closer to linear force-free field and suggested this to be an indicator of incomplete Taylor relaxation. Recently, \citet{2023L} found some evidence for Taylor relaxation in increased homogenization of $\alpha$ for multiple X-class flares during 2010-2017.

Along with the observational studies, numerical simulations employing analytical magnetic fields have also been carried out. The simulation by \citet{2000A} employed bipolar potential fields driven by a 2D velocity field imposed at the bottom boundary. They found the terminal state of their simulation to be far from a constant-$\alpha$ field. Contrarily, \citet{2008Br} and \citet{2009Ho} investigated the nanoflare heating model by following the development of  kink instability in coronal loops. The relaxed state was found to be consistent with a linear force-free configuration. In the context of topological dissipation problem (\citealp{TDP}), \citet{2011P} used braided magnetic fields and found the terminal state of relaxation to be nearly nonlinear force-free. For resistive MHD simulation of a solar coronal jet, \citet{2015P} analyzed the evolution of helicity for several gauge choices, and found it to be approximately conserved. Recently, \citet{2023R} explored the formation of a magnetic flux rope in MHD simulation of the Quiet Sun, where disordered low-lying coronal magnetic field lines undergo multiple small-scale reconnections. The authors recognized the process as self-organization, where an inverse cascade of helicity occurs, making the system to tend toward Taylor relaxation. 

The above simulations although explore various elements of relaxation in coronal transients but are idealized scenarios because of their well-organized analytical initial conditions. In this paper a novel approach to study the magnetic relaxation is adopted where the initial field of a flaring region is obtained through extrapolation of coronal field using photospheric magnetograms. The idea is to accommodate the field line complexity of an actual flaring region which may not get captured in analytical magnetic fields. Subsequently, MHD simulation in combination with multiwavelength analyses is carried out to explore the reconnection dynamics and its consequence on the magnetic relaxation.

The paper is organized as follows. Section \ref{sec2} describes the active region, the M1.3 class flare and the rationale behind selecting the AR. It also describes the temporal development of the transient activity using Extreme Ultraviolet (EUV) observations from SDO/AIA satellite. In Section \ref{sec3}, we present the details and setup for magnetic field extrapolation, supplemented with morphological investigation and indices that characterize the accuracy of magnetic field reconstruction. In Section \ref{sec4}, the numerical framework for magnetohydrodynamics simulation is presented. Section \ref{sec5} presents the results and analysis of this study and in Section \ref{sec6}, we present the summary and discussion of this work. 

\section{Active Region and Flare}
\label{sec2}

\begin{figure}[H]
\centering
\includegraphics[scale=0.8]{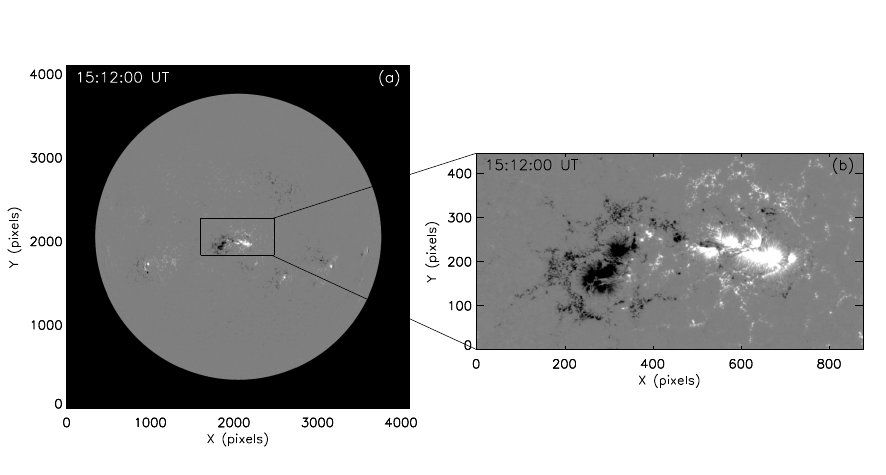}
\caption{(a) The line-of-sight magnetogram at 15:12 UT on 04 January, 2015, from SDO/HMI. The corresponding dimensions in pixel units are 4096$\times$4096 (b) CEA projected and cropped magnetogram (based on HARP active region patch) at 15:12 UT, having dimensions 877$\times$ 445 in pixel units for AR NOAA 12253. Both the figures are scaled to represent magnetic field strength within $\pm$\,1000 Gauss, with \textit{black} patches representing the negative polarity and \textit{white} patches representing the positive polarity.}
\label{fig1}
\end{figure}

The choice of active region (AR) and solar flare in this study is governed by the following rationale (a) The flare should be GOES (Geostationary Operational Environmental Satellite) M class or higher to ensure significant dissipation in magnetic energy (b) In the post-flare phase, there should not be any other major flaring activity so that the magnetic energy buildup and decay phases are sharp and clear (c) Due to the use of magnetic field extrapolation model in this study, the chosen AR should be nearly disk centered. Such a choice pertains to low measurement error in photospheric vector magnetic field and also minimizes the projection effects due to finite curvature of the photospheric surface (e.g. see \citealp{1988V}). Both the effects combinedly reduce error during magnetic field extrapolation. Considering these constraints, we select the AR NOAA 12253, with heliographic coordinates as S05E01 on January 4, 2015. It hosts a GOES M1.3 class flare of net duration 35 minutes (min.), having start, peak, and end time as 15:18 UT, 15:36 UT, and 15:53 UT, respectively. Importantly, in the post-flare phase, there is no flaring activity for the next six hours. Along with the aforementioned criterion's, we make sure that the selected active region complies with the condition $\mathrm{B_{z}}$ = const. at the bottom boundary, used in the MHD simulation. This translates into the requirement that during the course of flaring activity, the total relative change in magnetic flux (integrated over the bottom boundary) is minimal. We use the line-of-sight magnetograms from hmi.M\_\,45 series of the Helioseismic Magnetic Imager (SDO/HMI: \citealp{2012Schou}, \citealp{2012Scherrer}) onboard the Solar Dynamics Observatory (SDO: \citealp{2012P}), with temporal cadence of 45 seconds to evaluate this. The original magnetogram (Panel (a) in Figure \ref{fig1}) having dimensions of 4096$\times$4096 in pixel units is CEA projected (\citealp{2002cg}) and cropped to match the pre-defined dimensions of HARP active region patch (\citealp{2014H}) for AR NOAA 12253, which is 877$\times$445 in pixel units (Panel (b) in Figure \ref{fig1}). Using this processed magnetogram, we find that over a period of 72 min., starting from 15:00 UT up to 16:12 UT, the relative changes in positive and negative flux with respect to their initial values are 0.36\,\% and 0.42\,\%, respectively.

\begin{figure}[H]
\centering
\includegraphics[scale=0.3]{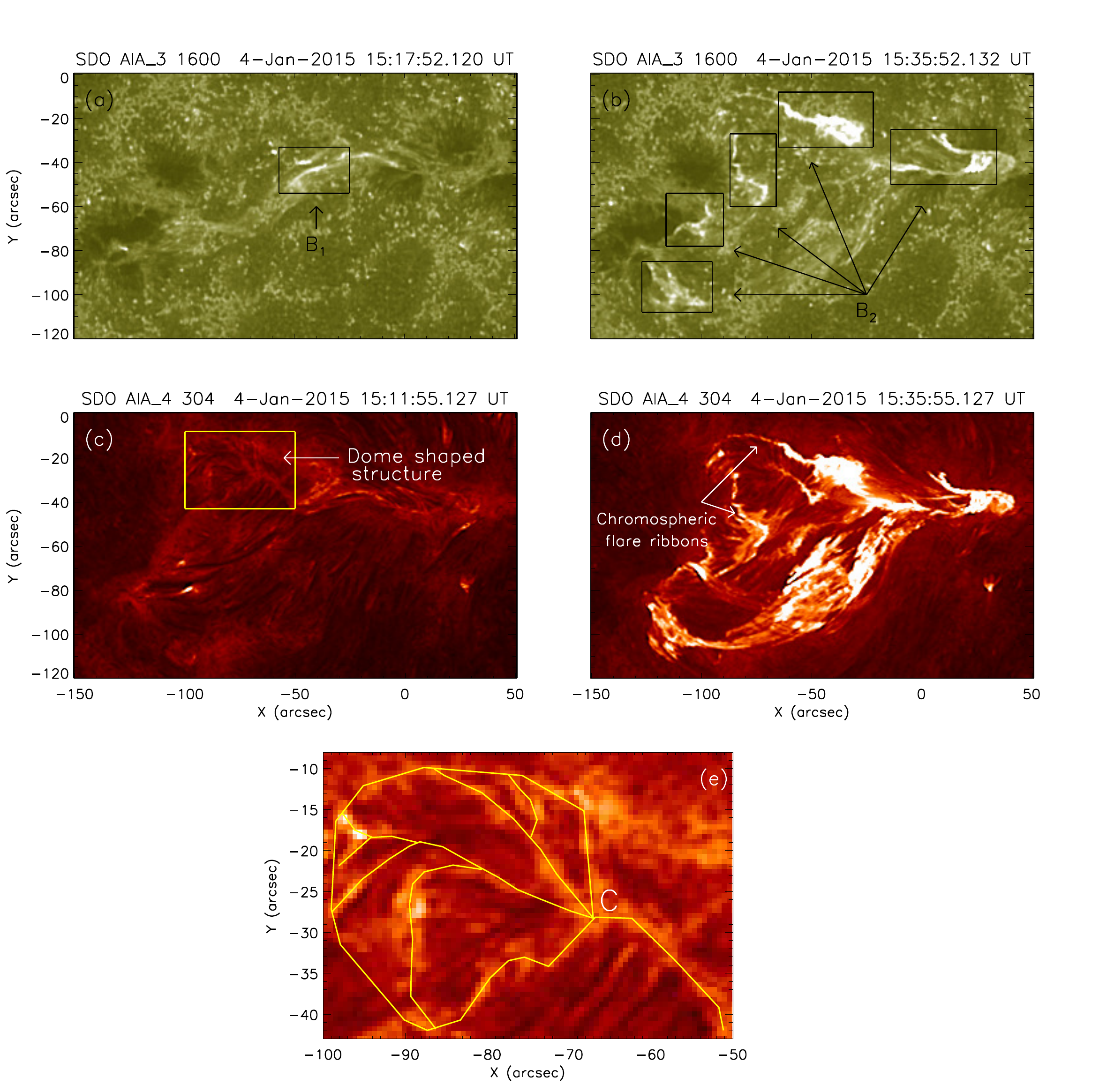}
\caption{Snapshots from observations of the solar flare in 1600\,\AA\, and 304\,\AA\, channels of SDO/AIA. Panels (a) and (b) reveal the brightening locations in 1600\,\AA\, during the beginning and peak phases of flare, marked by $\rm{B_{1}}$ and $\rm{B_{2}}$. Panels (c) and (d) highlight the initial configuration of the identified dome-shaped structure and chromospheric flare ribbons during the peak phase of flare. A zoomed in image of the boxed region in Panel (c) is presented in Panel (e) with enhanced image contrast. The \textit{yellow} color lines represent the manual tracing of the structure, while C labels the central location, where all the lines meet (The spatiotemporal evolution of flare in 1600\,\AA\, and 304\,\AA\, channels is available as high resolution movie in supplementary material).} 
\label{fig2}
\end{figure}

\noindent We explore the spatiotemporal evolution of flare using observations from 1600\,\AA\, and 304\,\AA\, channels of the Atmospheric Imaging Assembly (SDO/AIA: \citealp{2012L}). In Figure \ref{fig2}, Panel (a) depicts the location of a brightening (labeled $\rm{B_{1}}$) during the beginning of flare. The location is relevant because it might host a potential reconnection site and merits attention. The subsequent evolution reveals multiple brightenings during the flare peak (labeled $\rm{B_{2}}$), as shown in Panel (b). Notably, in the observations of 304\,\AA\, channel, we find the presence of a dome shaped structure. Its spatial location is marked by the \textit{yellow} colored box in Panel (c). A zoomed in view of this boxed region, with better image contrast, is given in Panel (e). The Panel highlights the approximate edges of the structure by \textit{yellow} lines. These lines depict multiple connections between the central location C and the traced, nearly circular periphery. Furthermore, the line toward the west of C indicates the association of dome structure with magnetic morphology in rest of the active region. We note that the overlaid lines are in agreement with the expected two dimensional projection of a dome structure. Further developments reveal the complete spatial extent of flare dynamics where specific chromospheric flare ribbons are recognizable, as marked in Panel (d) by \textit{white} arrows. Therefore, in observations, the brightenings $\rm{B_{1}}$, $\rm{B_{2}}$ and the dome shaped structure are identified to be of significance and hence, merit investigation of associated magnetic field line morphologies for an understanding of their role in the flaring activity and relaxation process.

\section{Magnetic Field Extrapolation}
\label{sec3}

In order to explore the magnetic field line morphology of the active region, we employ a non-force-free extrapolation model (\citealp{Ramit2004}, \citealp{Ramit2007} and \citealp{Hu2010}). The rationale for using the non-force-free extrapolation follows from an order of magnitude estimate for Lorentz force and rate of change of momentum on the photosphere (see \cite{2022Ag} for details). The estimate is obtained by using equation (\ref{lorentz}), as follows
\begin{equation}
\frac{\mid\bf{J}\times\bf{B}\mid}{\Big{|}\rho\frac{dv}{dt}\Big{|}}\sim\frac{1}{\beta},
\label{lorentz}
\end{equation}
where all the quantities have usual meanings. Since, $\beta \approx 1$ on the photosphere, equation (\ref{lorentz}) then gives
\begin{equation}
|\mathbf{J}\times\mathbf{B}|\sim\Big{|}\rho\frac{dv}{dt}\Big{|},
\end{equation}
thus making Lorentz force a plausible driver for the photospheric motions and an apt candidate to initiate MHD simulations. The NFFF extrapolation exploits a magnetic field $\mathbf{B}$ satisfying an inhomogeneous double curl Beltrami equation \citep{Ramit2007}
\begin{equation}
\nabla\times(\nabla\times\mathbf{B})+a\nabla\times\mathbf{B}+b\mathbf{B}=\nabla \psi
\label{e:bnff},
\end{equation}
having $a$ and $b$ as constants. The solenoidality of ${\bf{B}}$ imposes $\nabla^2\psi=0$. Notably, the double-curl equation represents a self-organized state satisfying the MDR principle, \citep[see][and references therein for details]{{Ramit2004}}. To solve the double-curl equation, an auxiliary field ${\bf{B}}^\prime={\bf{B}}-(\nabla\psi)/{b}$ \citep{Hu2008}---satisfying the corresponding homogeneous equation is constructed. The equation represents a two-fluid steady state \citep{1998MY} and has a solution
\begin{equation}
\mathbf{B}^\prime = \sum_{i=1,2} \mathbf{B}_i
\label{e:b123}
\end{equation}
The $\mathbf{B}_{i}$ are Chandrasekhar-Kendall eigenfunctions \citep{1957SC}, obeying force-free equations
\begin{equation}
    \label{e:b124}
    \nabla\times{\bf{B}}_i=\alpha_i{\bf{B}}_i,
\end{equation}
with constant twists $\alpha_i$, and form a complete orthonormal set when the 
eigenvalues are real \citep{GY}. Straightforwardly, 
\begin{equation}
\mathbf{B} = \sum_{i=1,2} \mathbf{B}_i+\mathbf{B}_3,
\label{e:b125} 
\end{equation}
where $\mathbf{B}_3=(\nabla\psi)/b$ is a potential field. 
Combining equations (\ref{e:b124}) and (\ref{e:b125}), we have\begin{equation}\begin{pmatrix}\B_1\\ \B_2\\ \B_3\end{pmatrix}=\mathcal{V}^{-1}\begin{pmatrix}\B\\ \nabla\times\B\\ \nabla\times(\nabla\times\B)\end{pmatrix},\end{equation}where the matrix $\mathcal{V}$ is a Vandermonde matrix having elements $\alpha^{i-1}_j$ for $i, j = 1, 2, 3,$ and $\alpha_{3}=0$ \citep{Hu2008}. The double-curl is solved by using the technique described in \citet{Hu2010}. A pair of $\alpha_i$ are selected and $\mathbf{B}_3$ is set to $\mathbf{B}_3=0$. Using $B_z$ from the observed magnetogram, along with $\alpha_{i}$, the \textit{z}-components of $\mathbf{B}_1$ and $\mathbf{B}_2$ are obtained at the bottom boundary. Afterwards, a linear force-free solver is employed to extrapolate the transverse components of $\mathbf{B}_1$ and $\mathbf{B}_2$.  Subsequently, an optimal pair of $\alpha_i$ is obtained by minimizing the average normalized deviation of the magnetogram transverse field $(\mathbf{B}_t$) from its extrapolated value
$(\mathbf{b}_t=\mathbf{B}_{1t}+\mathbf{B}_{2t}$), quantified as
\begin{equation}
E_n =\sum_{i=1}^M |\mathbf{B}_{t,i}-\mathbf{b}_{t,i}|/\sum_{i=1}^M |\mathbf{B}_{t,i}|,
\end{equation}
where $M$=$N^2$ is the total number of grid-points on the transverse plane. The ${E_n}$ is further reduced by employing the following decomposition for $\mathbf{B}_{3}$ 
\begin{equation}
\mathbf{B}_{3} = \mathbf{B}_{3}^{(0)} + \mathbf{B}_{3}^{(1)} + \mathbf{B}_{3}^{(2)} + ........ + \mathbf{B}_{3}^{(k)},
\end{equation}
where $\mathbf{B}_{3}^{(0)}=0$. Now, using $\mathbf{b}_t=\mathbf{B}_{1t}+\mathbf{B}_{2t}+\mathbf{B}_{3t}^{(k)}$, the transverse difference $\triangle \mathbf{b}_{t}=\mathbf{B}_{t}-\mathbf{b}_{t}$ is obtained and further utilized to estimate the \textit{z}-component of $\mathbf{B}_{3}^{(k+1)}$. Subsequently, $\mathbf{B}_{3t}^{(k+1)}, \mathbf{b}_{t}$, $E_{n}$ are estimated, and the procedure is repeated until the value of  $E_n$ approximately saturates with the number of iterations, making the solution unique. Importantly, the procedure alters the bottom boundary and a correlation with the original magnetogram is necessary to check for the accuracy.

For our purpose, the vector magnetogram at 15:12 UT, from the hmi.sharp\_\,cea\_\,720s series (\citealp{2014bobra}) of SDO/HMI is employed as the bottom boundary. Though the SHARP series accounts for projection and foreshortening effects, a disk-centered active region reduces the possibility of any distortion in the magnetogram. The magnetic field components on the photosphere are obtained as $\rm{B_{r}},\,B_{p}$, and $\rm{B_{t}}$, which satisfy (a) $\rm{B_z}$=$\rm{B_{r}}$ (r; radial), (b) $\rm{B_x}$=$\rm{B_{p}}$ (p; poloidal), and (c) $\rm{B_y}$=$\rm{-B_{t}}$ (t; toroidal) in a Cartesian coordinate system. The dimensions of the observed magnetogram is 877$\times$445 pixels ($\approx$\,317.91 Mm$\times$161.31 Mm). To save computational cost, we suitably crop and scale the magnetogram to new dimensions of 216$\times$110 pixels ($\approx$\,313.2 Mm$\times$159.5 Mm) and extrapolation is carried out in a computational box defined by 216$\times$110$\times$110 voxels, where a voxel represents a value on a regular grid in 3-D space. The cropping and scaling procedures render the relative changes in positive and negative magnetic fluxes to be 0.02\,\% and 0.84\,\%, respectively. These changes are minimal and approximately satisfy the condition, $\mathrm{B_{z}}$ = const. in the MHD simulation.

\subsection{Robustness of Extrapolated Magnetic Field}
\label{s3.1}

The robustness of magnetic field extrapolation is quantified by computing the following parameters.\\

\noindent (a) The current weighted average ($\sigma_{j}$) of the sine of angle ($\sigma_{i}$) between current density and magnetic field (\citealp{Wheat2000}), as defined in equation (\ref{E1})
\begin{equation}
\sigma_{j} = \frac{\sum_{i}J_{i}\sigma_{i}}{\sum_{i}J_{i}},\sigma_{i} = \frac{|\mathbf{J}\times\B|_{i}}{J_{i}B_{i}}, \label{E1}
\end{equation}
where $i$ runs over all the voxels in the computational box. Afterwards, the sine inverse of $\sigma_{j}$ (denoted by $\theta$) is computed, which represents the average angle between current density and magnetic field. In our case, $\theta=63.73^{\circ}$, which is expected because the model is non-force-free at the bottom boundary.\\

\noindent (b) The fractional flux (\citealp{2020GIL}), which quantifies the divergence free condition of the magnetic field, as defined in equation (\ref{E2})
\begin{equation}
\langle |f_{d}| \rangle = \Bigg{\langle} \frac{\int_{\partial S_{i}}\B\cdot d\textbf{S}}{\int_{S_{i}}|\B|dV}\Bigg{\rangle}, \label{E2}
\end{equation}
where $\partial S_{i}$ represents the surface area of any voxel and $S_{i}$ it's volume. We find this value to be $3.366 \times 10^{-9}$, which is numerically small enough to justify the solenoidal property of extrapolated magnetic field.\\

\begin{figure}[H]
\centering
\includegraphics[scale=0.5]{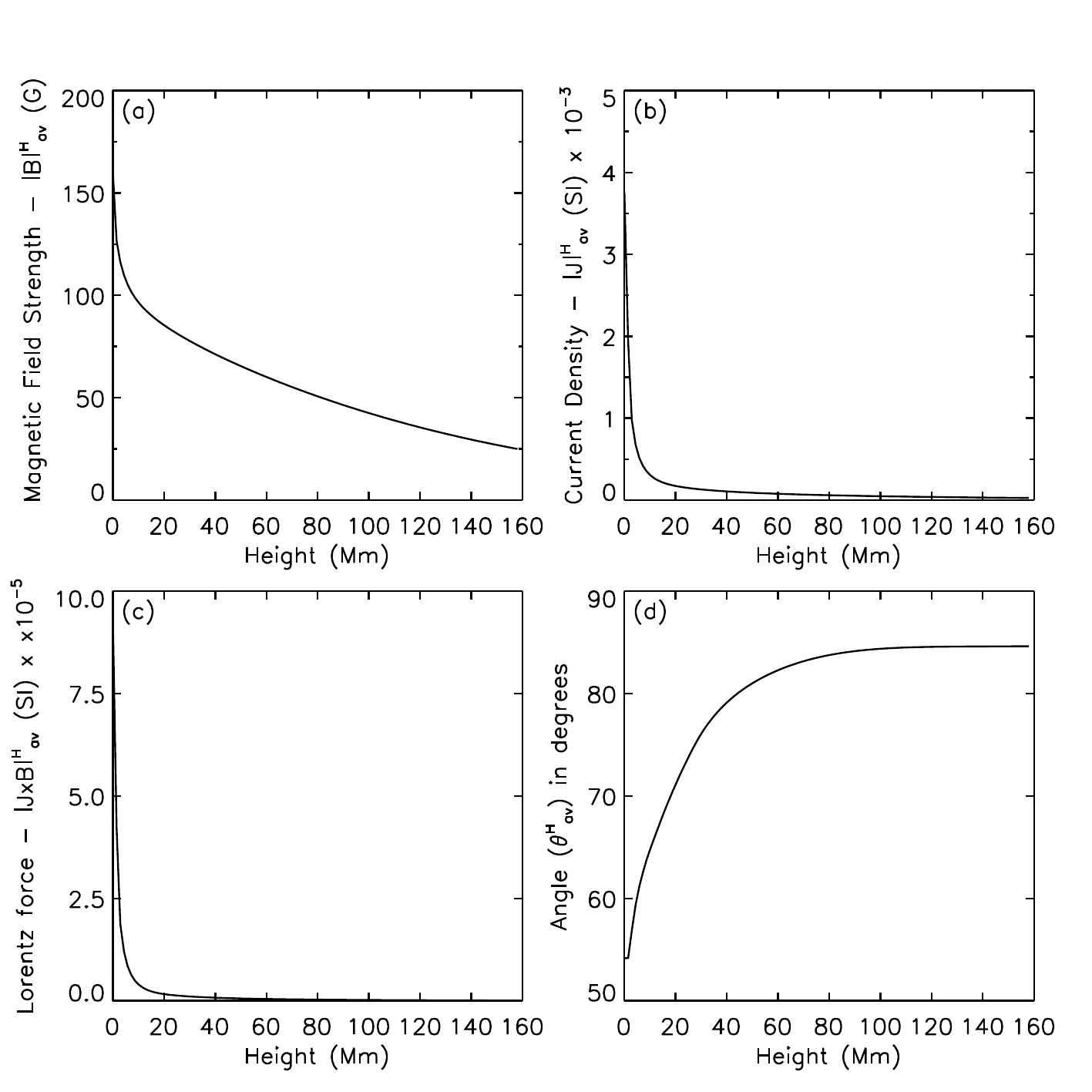}
\caption{Variation of horizontally averaged (a) magnetic field strength (b) current density (c) Lorentz force, and (d) $\theta$ with height for the extrapolated magnetic field.}
\label{fig3}
\end{figure}

\noindent (c) The ratio of total magnetic energy with respect to the total potential state energy, denoted by $E_{\mathrm{\rm{NFFF}}}/E_{\mathrm{P}}$. It allows to evaluate the free magnetic energy, hence sheds light on the capability of model to account for energy released during the transient phenomenon. The ratio turns out to be 1.305 in our case, hence, suggests that the extrapolated magnetic field has $\approx\,30.5\%$ more energy than the potential field. Quantitatively, the amount of available free energy is $5.6\times10^{31}$ ergs, which is enough to power a GOES M class flare. Further, to characterize the extrapolated magnetic field in the solar atmosphere, we check the variation of horizontally averaged magnetic field strength ($|\B|^{\rm H}_{\rm av}$), current density ($|\mathbf{J}|^{\rm H}_{\rm av}$), Lorentz force ($|\mathbf{J}\times\B|^{\rm H}_{\rm av}$), and $\theta^{\rm H}_{\rm av}$ with height, as shown in Figure \ref{fig3}. The averages are defined as

\begin{align}
&{|\B|}^{\rm H}_{\rm av} = \frac{1}{N}\times\displaystyle\sum_{l=0}^{N-1}\sqrt{{|\B|}_{l}^{2}},\\
&{|\mathbf{J}|}^{\rm H}_{\rm av} = \frac{1}{N}\times\displaystyle\sum_{l=0}^{N-1}\sqrt{{|\mathbf{J}|}_{l}^{2}},\\
&{|\mathbf{J}\times\B|}^{\rm H}_{\rm av}=\frac{1}{N}\times\displaystyle\sum_{l=0}^{N-1}\sqrt{{|\mathbf{J}\times\B|}_{l}^{2}},\\
&{\theta}^{\rm H}_{\rm av} = \frac{\displaystyle\sum_{l=0}^{N-1}\left[\sqrt{{|\mathbf{J}|}_{l}^{2}}\times{\sigma}_{l}\right]}{\displaystyle\sum_{l=0}^{N-1}\sqrt{{|\mathbf{J}|}_{l}^{2}}}, \sigma_{l}=\frac{\displaystyle\sqrt{{|\mathbf{J}\times\B|}_{l}^{2}}}{\displaystyle\sqrt{{|\mathbf{J}|}_{l}^{2}}\times\sqrt{{|\B|}_{l}^{2}}},
\end{align}    
where $l$ denotes the voxel index. The averages are computed over layers defined by the 2D arrangement of voxels having $N = 216\times 110$ voxels along the $x-$ and $y-$directions, respectively. Panel (a) reveals a continuous decrease of magnetic field strength with height and with respect to the bottom boundary, the percentage decrement is $84.4\,\%$. The profiles in Panels (b) and (c) highlight the rapid decay of respective quantities, which approach saturation asymptotically. Within a distance of nearly 3\,Mm, both the quantities decay by almost $50$\,\%, but the subsequent decrease is relatively slower. In the higher layers of solar atmosphere, magnetic field lines tend to be more potential, thus characterized analytically by zero current density and low twist. Therefore, as shown in Panel (d), $\theta$ tends to increase with height. 

\subsection{Morphological Investigation of Magnetic Field Lines}

\begin{figure}[H]
\centering
\includegraphics[scale=0.5]{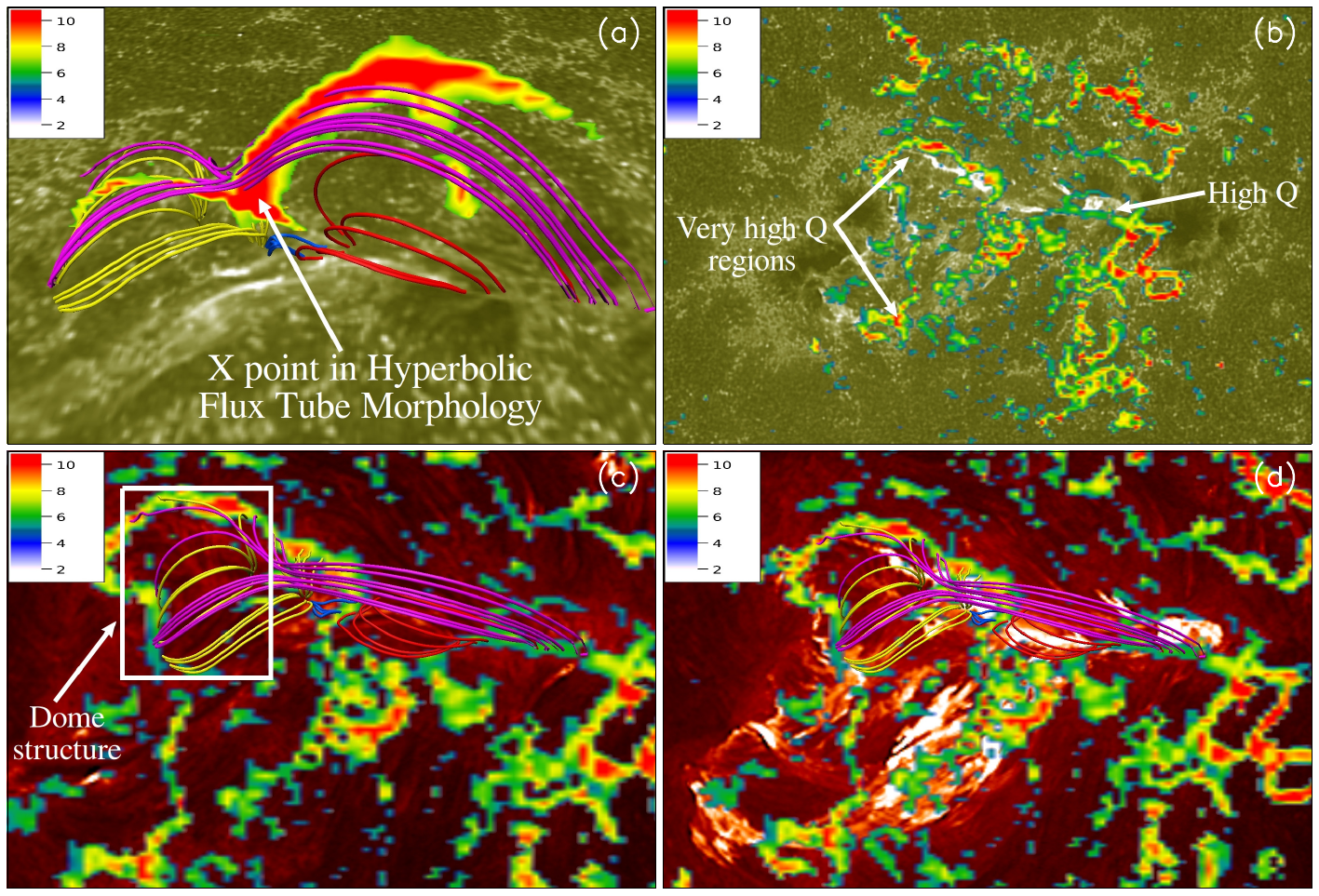}
\caption{Regions of high squashing degree and morphology of the hyperbolic flux tube (HFT). Panels (a), (b) and Panels (c), (d) are overlaid with observations in 1600\,\AA\, and 304\,\AA \, channel of SDO/AIA at the bottom boundary, corresponding to the same time instants as in Figure \ref{fig2}. In all the Panels, the map of squashing degree ln\,\textit{Q} is given with color table. In Panel (a), the map is perpendicular to the bottom boundary, crossing through the HFT, while in Panels (b), (c) and (d), the ln\,\textit{Q} map is in the plane of the bottom boundary. Panel (a) shows HFT from a side view, while Panels (c) and (d) show HFT from a top-down view. Panels (a) and (c) use a zoomed in viewpoint while Panels (b) and (d) use a zoomed out viewpoint. In Panel (c), the dome shaped structure is marked with a \textit{white} color box.}
\label{fig4}
\end{figure}

Focusing on brightenings $\rm{B_{1}}$, $\rm{B_{2}}$, and the dome structure, magnetic field line morphologies in the active region are explored. Cospatial with $\rm{B_{1}}$, a hyperbolic flux tube (HFT; \citealp{2002TH}) is found, as shown in Figure \ref{fig4}. Panel (a) of the figure shows magnetic field line linkage of the HFT configuration --- constituted by the four quasi-connectivity domains (\citealp{2014Zhao}) in \textit{blue}, \textit{yellow}, \textit{pink}, and \textit{red} colors. These domains comprise of two intersecting quasi separatrix layers or QSLs (\citealp{2002TH}) --- one by \textit{blue} and \textit{yellow} magnetic field lines (MFLs), the other by \textit{pink} and \textit{red} MFLs. Notably, these configurations are preferred sites for reconnection (\citealp{2006Demoulin}) and hence are of interest. QSLs are characterized by strong but finite gradients in the magnetic field line mapping. The gradients are quantified by the estimation of squashing degree $Q$, defined as follows. For the two footpoints of a field line, rooted in  $\vec{R}_{a}(x_{a},y_{a})$ and $\vec{R}_{b}(x_{b},y_{b})$, the Jacobian matrix for the mapping $\prod_{ab}$: $\vec{R}_{a}(x_{a},y_{a}) \mapsto \vec{R}_{b}(x_{b},y_{b})$ is given by
\begin{equation}
D_{ab}=\left[\frac{\partial\vec{R}_{a}}{\partial\vec{R}_{b}}\right] = 
\begin{pmatrix}
\partial x_{b}/x_{a} & \partial x_{b}/y_{a} \\
\partial y_{b}/x_{a} & \partial y_{b}/y_{a} \\
\end{pmatrix} = \begin{pmatrix}
a & b\\
c & d
\end{pmatrix},
\end{equation}
from which, the squashing degree $Q$ is defined as
\begin{equation}
Q \equiv \frac{a^{2}+b^{2}+c^{2}+d^{2}}{|B_{n,a}(x_{a},y_{a})/B_{n,b}(x_{b},y_{b})|},
\end{equation}
where $B_{n,a}(x_{a},y_{a})$ and $B_{n,b}(x_{b},y_{b})$ are the normal components of magnetic field at the respective footpoints. We calculated the squashing degree by using the numerical code developed in \citet{Liu2016}, available at \href{http://staff.ustc.edu.cn/~rliu/qfactor.html}{http://staff.ustc.edu.cn/$\sim$rliu/qfactor.html}. In presence of strong gradients (high $Q$ value), magnetic field lines undergo slippage while passing through the current layers, often referred as slipping reconnection (\citealp{2006Au}). Panel (a) shows the ln\,\textit{Q} map in a plane perpendicular to the bottom boundary, and crossing through the HFT morphology. It reveals the characteristic X-shape (ln\,$Q$\,$\geq$\,8) along HFT, which further confirms our interpretation of the morphology. Similarly, in Panel (b), regions of high gradient in the plane of the bottom boundary are seen to be nearly cospatial with $\rm{B_{2}}$, thus suggesting a plausible scenario for slipping reconnection.

\noindent In particular, as evident from Panels (c) and (d), the footpoints of \textit{yellow} MFLs lie on the boundary of dome, while those at one end of \textit{pink} MFLs partially cover the periphery of dome. The presence of high gradients in footpoint mapping of the MFLs is indicative of slippage, which can possibly explain parts of brightening $\rm{B_{2}}$ and chromospheric flare ribbons. The robustness of the extrapolated magnetic field and it's agreement with observations suggests that it can be reliably utilized as an input for the reported magnetohydrodynamics simulation.

\section{Magnetohydrodynamics Simulation}
\label{sec4}

To successfully simulate the coronal transients, the condition of flux-freezing must hold everywhere in the computational box, while allowing for magnetic reconnection at the plausible locations. In this work, the coronal plasma is idealized to be thermodynamically inactive and incompressible. The governing MHD equations in dimensionless form are

\begin{eqnarray}
\derivp{\vec v}{t} + (\vec v \cdot \nabla)\vec v &=& -\nabla p + (\nabla \times \vec B) \times \vec B + \frac{1}{R_F^A}\nabla^{2}\vec v, \label{eqn1} \\
\derivp{\vec B}{t} &=& \nabla \times (\vec v \times \vec B), \label{eqn4}\\
\nabla \cdot \vec v &=& 0, \label{eqn3} \\
\nabla \cdot \vec B &=& 0, \label{eqn2}
\end{eqnarray}

\noindent where $R_F^A=(V_{A}L)/\nu$ is an effective fluid Reynolds number with $V_{A}$ as the Alfv\'en speed and $\nu$ as the kinematic viscosity. Hereafter, $R_F^A$ is referred as fluid Reynolds number to keep the terminology uncluttered. The dimensionless equations are obtained by the normalization listed below.

\begin{equation}
\vec B \rightarrow \frac{\vec B}{B_{0}}, \vec v \rightarrow \frac{\vec v }{V_{A}}, L \rightarrow \frac{L}{L_{0}}, t \rightarrow \frac{t}{\tau_{a}}, p \rightarrow \frac{p}{\rho_{0} V_{A}^{2}}.
\end{equation}

\noindent In general, $B_{0}$ and $L_{0}$ are characteristic values of the system under consideration. Importantly, although restrictive, the incompressibility may not affect magnetic reconnection directly and has been used in earlier works (\citealp{1991Dahl}, \citealp{2005Auli}). Moreover, utilizing the discretized incompressibility constraint, the pressure $p$ satisfies an elliptic boundary value problem on the discrete integral form of the momentum equation. 

Toward simulating relaxation physics, it is desirable to preserve flux-freezing to an appropriate fidelity by minimizing numerical diffusion and dispersion errors away from the reconnection sites. Such minimization is a signature of a class of inherently nonlinear high resolution transport methods that prevent field extrema along flow trajectories while ensuring higher order accuracy away from the steep gradients in advected fields \citep{RBCLOW}. Consequently, equations (\ref{eqn1})-(\ref{eqn2}) are solved by the numerical model EULAG-MHD (\citealp{2013SC}), central to which is the spatio-temporally second order accurate, nonoscillatory, and forward-in-time Multidimensional Positive-Definite Advection Transport Algorithm, MPDATA (\citealp{smolar1983}, \citealp{Piotr1998}, \citealp{piotrsingle}). For the  computations carried out in the paper, important is the widely documented dissipative property of MPDATA that mimics the action of explicit subgrid-scale turbulence models \citep{Margo2001,Margo2002,Margo2006}, wherever the concerned advective field is under resolved---the property referred to as Implicit Large-Eddy Simulations (ILES; \citealp{Margolin2006,Piotr2007,gmr2010}). Therefore, the effective numerical implementation of the induction equation by EULAG-MHD is
\begin{equation}
    \derivp{\vec B}{t} = \nabla \times (\vec v \times \vec B) + {\bf{D_B}}, \label{new}
\end{equation}
where, $\bf{D_{B}}$ represents the numerical magnetic diffusion---rendering magnetic reconnections to be solely numerically assisted. 
Such delegation of the entire magnetic diffusivity to ILES is advantageous but also calls for a cautious approach in analyzing and extracting simulation results. Being localized and intermittent, the magnetic reconnection in the spirit of ILES minimizes the computational effort, while tending to maximize the effective Reynolds number of simulations (\citealp{2008JFM}). However, the absence of physical diffusivity makes it impossible to accurately identify the relation between electric field and current density—--rendering a precise estimation of magnetic Reynolds number unfeasible. Being intermittent in space and time, quantification of this numerical dissipation is strictly meaningful only in the spectral space where, analogous to the eddy viscosity of explicit subgrid-scale models for turbulent flows, it only acts on the shortest modes admissible on the grid \citep{2003DS}, particularly near steep gradients in simulated fields. Such a calculation is beyond the scope of this paper. Notably, earlier works (\citealp{Av2020}, \citealp{2022Y}, \citealp{Bora2022}, \citealp{2022Ag}) have shown the reconnections in the spirit of ILES to be consistent with the source region dynamics of coronal transients and provide the credence for adopting the same methodology here. 

\subsection{Numerical Setup}
\label{sec41}
The MHD simulation is carried out with bottom boundary satisfying the line-tied condition (\citealp{Au2010}). In our case, we ensure this by keeping $\rm{B_{z}}$ and $\rm{v_{z}}$ fixed at the bottom boundary. We have kept the lateral and top boundaries of the computational box open. The simulation is initiated from a static state (zero flow) using the extrapolated non-force-free magnetic field, having dimensions 216$\times$110$\times$110 which is mapped on a computational grid of $x$\,$\in$\ [-0.981,0.981], $y$\,$\in$\,[-0.5,0.5], and $z$\,$\in$\,[-0.5,0.5], in a  Cartesian coordinate system. The spatial step sizes are $\Delta$\,$x$\,=\,$\Delta$\,$y$\,=\,$\Delta$\,$z$\,$\approx$\,0.0091 ($\equiv$\,1450 km), while the time step is  $\Delta$\,$t$\,=\,2$\times$10$^{-4}$ ($\equiv$\,0.2544 sec). Using typical values in the solar corona, $L_{\rm{cor.}}$\,=\,100 Mm, $V_{A}|_{\rm{cor.}}$\,=\,100 \,$\mathrm{km\,s^{-1}}$, and $\nu_{\rm{cor.}}$\,=\,4$\times$10$^{9}$\,$\mathrm{m^{2}\,s^{-1}}$, the corresponding fluid Reynolds number is estimated to be, $R_F^A|_{\rm{cor.}}$\,=\,25,000. However, in the numerical setup for our simulation, $R_F^A|_{\rm{sim.}}$\,=\,5000$\equiv$\,0.2\,$\times$\,$ R_F^A|_{\rm{cor.}}$. This reduction in fluid Reynolds number may be envisaged as a smaller Alfv\'en speed, which turns out to be $V_A|_{\rm sim.}$\,$\approx$\,0.125$\times$\,$V_A|_{\rm cor.}$ for $L_{\rm{sim.}}$\,=\,110\,$\times$\,1450\,km\,=\,159.5 Mm. The total simulation time in physical units is equivalent to $nt\times\Delta t\times(L_{\rm{sim.}}/V_{A}|_{\rm{sim.}})\approx63.6$ min., where $nt$\ = \,15000. Importantly, although the coronal plasma with a reduced fluid Reynolds number is not realistic, the choice does not affect the changes in field line connectivity because of reconnection, but only the rate of evolution. Additionally, it saves computational cost, as demonstrated by \citet{Ji2016}.

\section{Results and Analysis}
\label{sec5}

The initial non-zero Lorentz force pushes the magnetofluid and generates dynamics. The overall simulated dynamics pertaining to magnetic relaxation can be explored from the evolution of the following grid averaged parameters

\begin{align}
&\mathrm{{W}^{V}_{av}} = \frac{1}{N}\times\displaystyle\sum_{l=0}^{N-1}{|\B|}_{l}^{2} \label{nd1}, \\
&\mathrm{{|\Gamma|}^{V}_{av}} = \frac{1}{N}\times\displaystyle\sum_{l=0}^{N-1} \bigg{|} (\mathbf{J}\cdot\B)_{l}\bigg{/}{|\B|}_{l}^{2}\bigg{|} \label{nd2}, \\
&\mathrm{{(J/B)}^{V}_{av}} = \frac{1}{N}\times\displaystyle\sum_{l=0}^{N-1}\sqrt{\frac{{|\mathbf{J}|}_{l}^{2}}{{|\B|}_{l}^{2}}} \label{nd3},
\end{align}
\noindent where, $l$ denotes the voxel index and the volume $\mathrm{V}$
encloses the volume of interest. $\mathrm{{W}^{V}_{av}}$ and $\mathrm{{|\Gamma|}^{V}_{av}}$ measure the grid averaged magnetic energy and twist, whereas $\rm{(J/B)}^{V}_{av}$ serves as a proxy to quantify gradient of magnetic field.

The plot of grid averaged magnetic energy, depicted in Panel (a), 
shows a continuous decrease ($\approx 7\%$) and is in alignment with the 
possibility of magnetic relaxation through reconnection. To support this 
idea, Panel (b) plots  the grid averaged twist $\mathrm{{|\Gamma|}^{V}_{av}}$. Notably, $\Gamma$ is also associated with magnetic helicity. The plot shows a decay of the average twist up to $\approx$ 40 minutes, followed by a rise. The initial decay is in conformity with the scenario of magnetic reconnection being responsible for untwisting of global field structure \citep{2010WS} and reducing the complexity of field lines. The scenario of reconnection assisted relaxation is further reinforced by a similar variation of $\rm{(J/B)}^{V}_{av}$ (Panel (c)) since, reconnection is expected to smooth out steep field gradients. The rise in both the parameters is due to a current enhancement localized near the top of the computational domain---addressed later in the paper. 

In the above backdrop, understanding dynamics of magnetic field lines 
involved in reconnection merits further attention. For the purpose, the computational volume is partitioned into three sub-volumes, as described in Figure \ref{t1}.

\begin{figure}[H]
\centering
\includegraphics[scale=0.12]{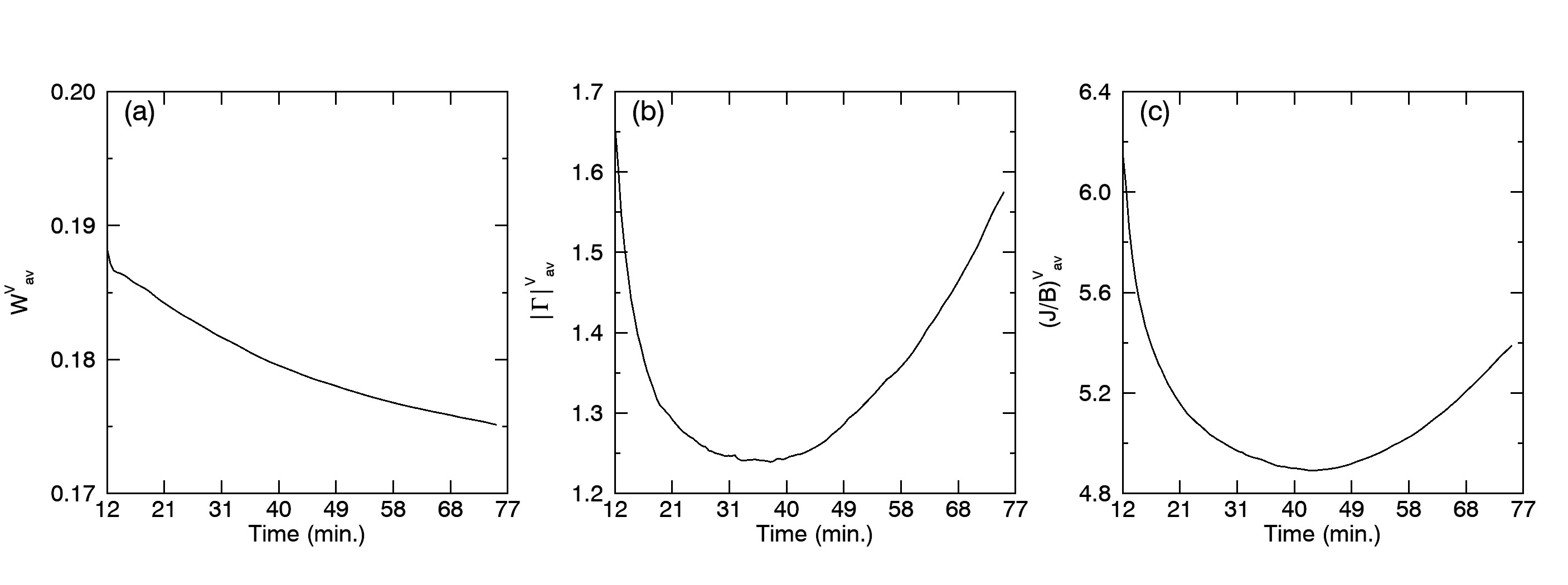}
\caption{Temporal evolution of grid averaged parameters (a) Magnetic energy ($\rm{W}^{V}_{av}$) (b) Twist ($\rm{|\Gamma|}^{V}_{av}$) (c) Gradient in magnetic field ($\rm{(J/B)}^{V}_{av}$). The origin of time scale maps from 15:12 UT.}
\label{fig5}
\end{figure} 

\begin{table}[ht]
\caption{The offset and extent (in voxels) for sub-volumes of interest ($\rm{S_{i}}$, i=1,2,3) in $x$-, $y$-, and $z$-directions}
\label{t1}
\begin{tabular}{ccccccc}
\specialrule{.1em}{.05em}{.05em}
 & $\rm{X_{off}}$ & $\rm{Y_{off}}$ & $\rm{Z_{off}}$ & $\rm{X_{size}}$ & $\rm{Y_{size}}$ & $\rm{Z_{size}}$ \\
\specialrule{.1em}{.05em}{.05em}
$\rm{S_{1}}$ & 102 & 56 & 0 & 8 & 10 & 5 \\
$\rm{S_{2}}$ & 80 & 50 & 0 & 60 & 30 & 20 \\
$\rm{S_{3}}$ & 70 & 20 & 0 & 70 & 60 & 110 \\
\specialrule{.1em}{.05em}{.05em}
\end{tabular}
\end{table}

\begin{figure}[H]
\centering
\includegraphics[scale=0.5]{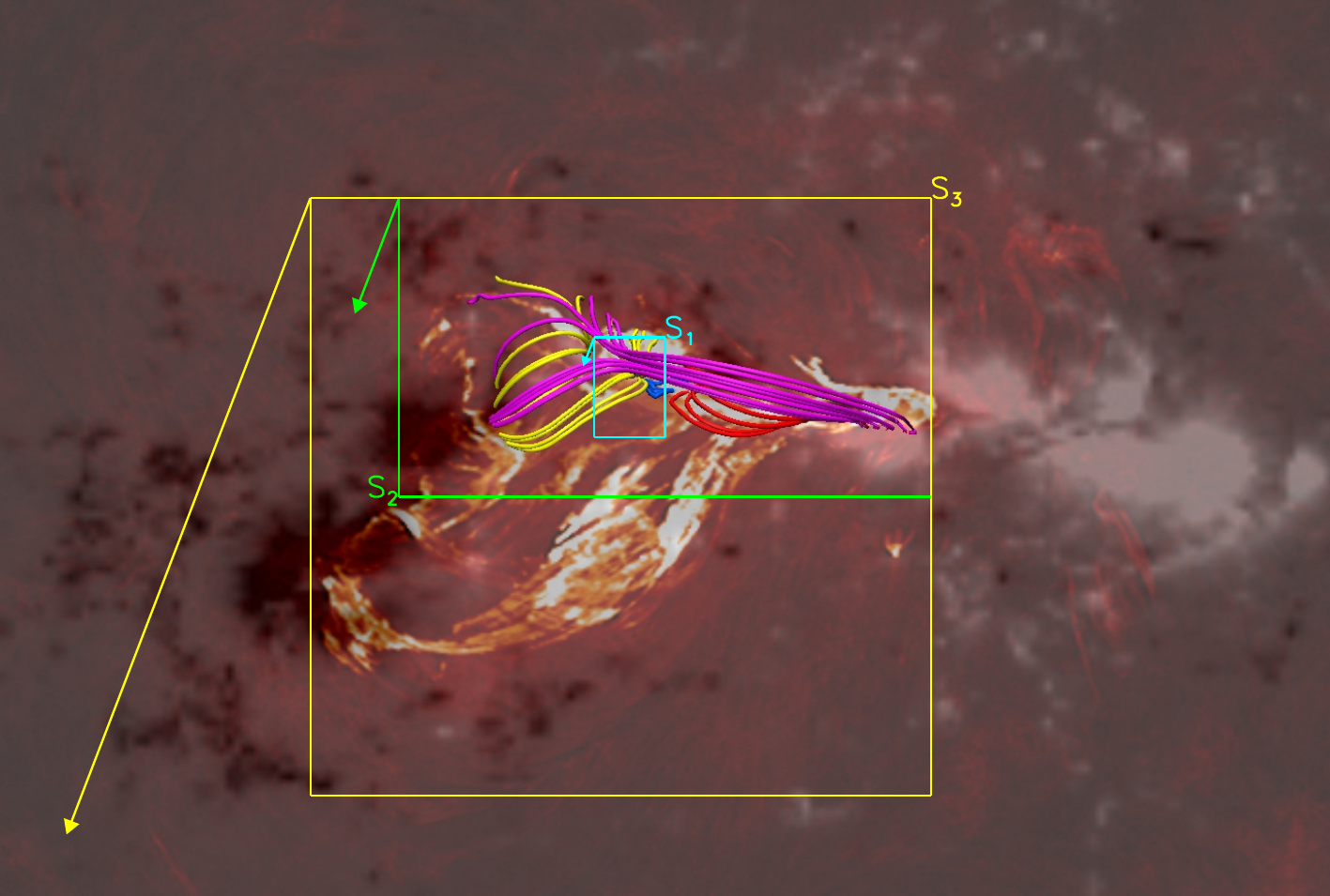}
\caption{Visual representation of sub-volumes $\rm{S_{1}}$, $\rm{S_{2}}$, and $\rm{S_{3}}$. The hyperbolic flux tube (HFT) configuration, overlaid with the vertical component of magnetic field and observation of the flaring event in 304\,\AA\, channel of SDO/AIA at 15:35:52 UT is shown. A zoomed in viewpoint is used, corresponding to a cutout of $150\times90$ pixels. The extent and spatial position of sub-volumes $\rm{S_{1}}$, $\rm{S_{2}}$, and $\rm{S_{3}}$ are marked by the \textit{cyan}, \textit{green}, and \textit{yellow} colored boxes. The arrows indicate the extent of sub-volumes along the $z$-direction and are drawn in proportion to the actual vertical sizes of sub-volumes given inFigure \ref{t1}.}
\label{fig6}
\end{figure}

\noindent Our selection focuses on the hyperbolic flux tube (HFT) as the principal reconnection site and on the observed extent of brightenings in the active region, as depicted in Figure \ref{fig6}. The two-dimensional projections of sub-volumes $\rm{S_{1}}$, $\rm{S_{2}}$, and $\rm{S_{3}}$ are shown in \textit{cyan}, \textit{green}, and \textit{yellow} color boxes. Sub-volume $\rm{S_{1}}$ enshrouds brightening $\rm{B_{1}}$ and is centered on the X-point of HFT, thus consisting of those regions where the development of strongest current layers is possible. $\rm{S_{2}}$ encloses the HFT morphology that envelops $\rm{B_{1}}$ and partly $\rm{B_{2}}$ such that the field line connectivities of depicted MFLs (see Figure \ref{fig4}) are contained within $\rm{S_{2}}$. Lastly, $\rm{S_{3}}$ covers the complete spatial extent of the observed brightening (see Figure \ref{fig2}) and full vertical height of the computational box.

\noindent Importantly, magnetic energy in a sub-volume can  change because of an interplay between dissipation, Poynting flux, and the conversion of kinetic energy to magnetic energy. Consequently, the following sections analyzes magnetofluid evolution for each sub-volume. For completeness, the analyses are augmented by plotting the corresponding grid averaged current densities:

\begin{equation}
\mathrm{{J}^{V}_{av}} = \frac{1}{N}\times\displaystyle\sum_{l=0}^{N-1}{|\mathbf{J}|}_{l}^{2} \label{nd4},     
\end{equation}
in usual notations. 

\subsection{Sub-volume $\rm{S_{1}}$}
\label{sec53}
\begin{figure}[H]
\centering
\includegraphics[scale=0.15]{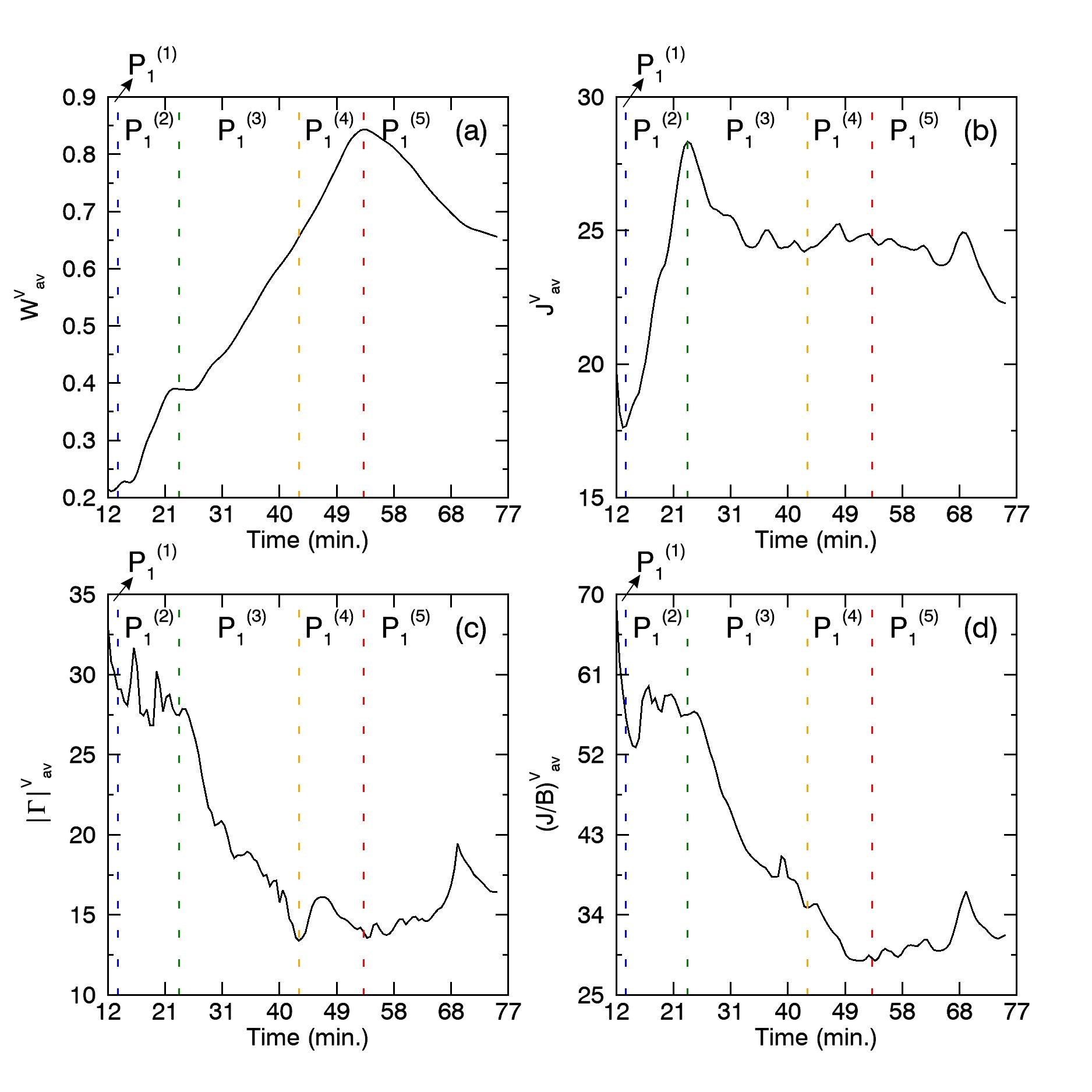}
\caption{Time evolution of grid averaged (a) magnetic energy ($\rm{W}^{V}_{av}$) (b) current density ($\rm{J}^{V}_{av}$) (c) twist parameter($\rm{|\Gamma|}^{V}_{av}$), and (d) $\rm{(J/B)}^{V}_{av}$ in sub-volume $\rm{S_{1}}$, during phases $\rm{P_{1}^{(1)}}$ (marked by the \textit{black} arrow), $\rm{P_{1}^{(2)}}$, $\rm{P_{1}^{(3)}}$, $\rm{P_{1}^{(4)}}$, and $\rm{P_{1}^{(5)}}$, respectively. The \textit{dashed} lines in \textit{blue}, \textit{green}, \textit{orange}, and \textit{red} colors separate the different phases in each of the profiles. The origin of the time scale maps to 15:12 UT.}
\label{fig7}
\end{figure}

The time evolution of  $\rm{W}^{V}_{av}$, $\rm{J}^{V}_{av}$, $\rm{|\Gamma|}^{V}_{av}$, and $\rm{(J/B)}^{V}_{av}$ is depicted in Panels (a), (b), (c), and (d) ofFigure \ref{fig7}, respectively. To understand their dynamical evolution, the time duration of numerical simulation is partitioned into five phases, denoted by $\mathrm{P_{1}^{(i)}}$, where, $\rm{i=1,2,...,5}$. The composition of $\rm{S_{1}}$ has five layers along the vertical direction, denoted by $\rm{z_{0}=0,1,...,4}$, and the contribution of each layer in the shaping of parameter profile needs to be examined.

$\rm{W}^{V}_{av}$ increases initially up to phase $\mathrm{P_{1}}^{(4)}$, followed by a continuous decay during $\mathrm{P_{1}}^{(5)}$. Auxiliary analysis (not shown here) reveals that the magnetic energy at all layers have a profile similar to $\rm{W}^{V}_{av}$. Contrarily, the same is not true for  $\rm{J}^{V}_{av}$ and $\rm{|\Gamma|}^{V}_{av}$---an explanation for which is presented below. 

Owing to $\rm{z_{0}=0,1}$, $\rm{J}^{V}_{av}$ decreases sharply in the beginning phase $\mathrm{P_{1}}^{(1)}$. During the rising phase $\mathrm{P_{1}}^{(2)}$, while all the layers exhibit similar profile, only $\rm{z_{0}=2,3,4}$ contribute most significantly because the X-point of HFT exists at these heights, which is a prominent site for development of strong currents. Lastly, from phase $\mathrm{P_{1}}^{(3)}$ to $\mathrm{P_{1}}^{(5)}$, there is an overall decrease in $\rm{J}^{V}_{av}$, again due to $\rm{z_{0}=2,3,4}$ along with some wiggling---predominantly due to $\rm{z_{0}=0,1,2}$. 

The $\rm{|\Gamma|}^{V}_{av}$ profile during phases $\mathrm{P_{1}}^{(1)}$ and $\mathrm{P_{1}}^{(2)}$ is shaped by $\mathrm{z_{0}=0,1}$. The decline during $\mathrm{P_{1}}^{(3)}$ is attributed to $\rm{z_{0}=2,3,4}$, while the evolution in $\mathrm{P_{1}}^{(4)}$ and $\mathrm{P_{1}}^{(5)}$ is strongly determined by the bottom two layers, i.e., $\rm{z_{0}=0,1}$. The pronounced effect of bottom two layers ($\rm{z_{0}=0,1}$) could be due to the fact that spatial structures near the bottom boundary are not sufficiently resolved due to reduction in resolution of the extrapolation. Consequently, the dynamics in near neighborhood of the X-point of HFT leads to fluctuations in the profile of $\rm{J}^{V}_{av}$ and $\rm{|\Gamma|}^{V}_{av}$, which do not smooth out due to the small size of sub-volume $\rm{S_{1}}$. 

Lastly, it is seen that the evolution of $\rm{(J/B)}^{V}_{av}$ is qualitatively similar to $\rm{|\Gamma|}^{V}_{av}$ profile. The quantitative changes in parameters during each of the phases are summarized inFigure \ref{table2} and from the rightmost column of this table, we note that the net magnetic energy and current density have increased while the overall twist and gradients have reduced in sub-volume $\rm{S_{1}}$. 

\begin{table}[h]
\caption{Summary of the quantitative changes in grid averaged profiles of magnetic energy ($\rm{W}^{V}_{av}$), current density ($\rm{J}^{V}_{av}$), twist parameter($\rm{|\Gamma|}^{V}_{av}$), and magnetic field gradient ($\rm{(J/B)}^{V}_{av}$) for sub-volume $\rm{S_{1}}$, during phases $\rm{P_{1}^{(1)}}$, $\rm{P_{1}^{(2)}}$, $\rm{P_{1}^{(3)}}$, $\rm{P_{1}^{(4)}}$, and $\rm{P_{1}^{(5)}}$, respectively. The positive and negative values indicate the rising and declining phases, while the net value in the rightmost column tells about the difference between terminal and initial states.}
\label{table2}
\begin{tabular}{ccccccc}     
\specialrule{.1em}{.05em}{.05em}
 & $\rm{P_{1}^{(1)}}$ & $\rm{P_{1}^{(2)}}$ & $\rm{P_{1}^{(3)}}$ & $\rm{P_{1}^{(4)}}$ & $\rm{P_{1}^{(5)}}$ & Net\\
\specialrule{.1em}{.05em}{.05em}
$\rm{W^{V}_{av}}$ & +0.004 & +0.170 & +0.266 & +0.188 & -0.188 & +0.440\\
$\rm{J}^{V}_{av}$ & -2.123 & +10.660 & -4.048 & +0.427 & -2.430 & +2.486\\
    $\rm{|\Gamma|}^{V}_{av}$ & -3.914 & -1.644 & -14.079 & +0.544 & +2.514 & -16.580\\
$\rm{(J/B)}^{V}_{av}$ & -12.957 & +0.392 & -21.705 & -5.667 & +2.598 & -37.340\\
\specialrule{.1em}{.05em}{.05em}
\end{tabular}
\end{table}

\begin{figure}[H]
\centering
\includegraphics[scale=0.18]{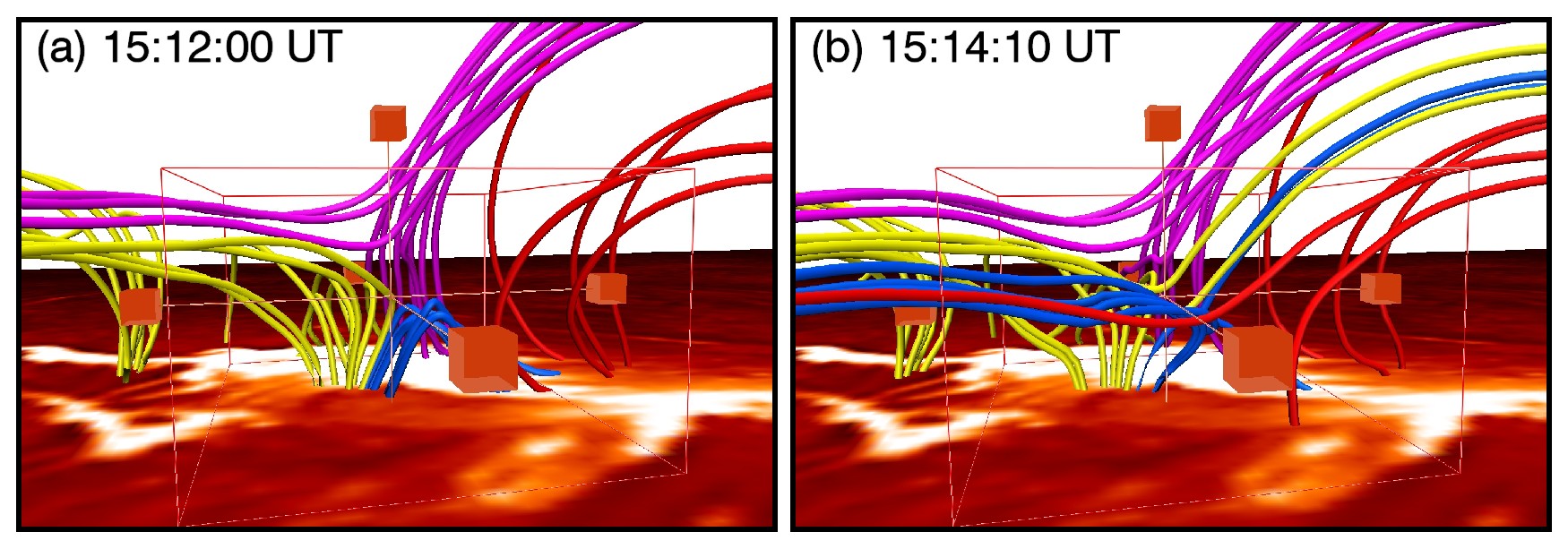}
\caption{Panels (a) and (b): Illustration of changes in the field line connectivity of \textit{yellow}, \textit{blue}, and \textit{red} MFLs due to reconnection at the X-point of HFT configuration. The \textit{red} colored box marks the edges of $\rm{S_{1}}$ while the bottom boundary is overlaid with an image in 304 \AA\, channel of SDO/AIA (The spatiotemporal evolution of the magnetic field line dynamics in sub-volume $\rm{S_{1}}$ is available as movie in supplementary materials).}
\label{fig8}
\end{figure}

\noindent The evolution of magnetic field line dynamics, as shown in Panels (a) and (b) ofFigure \ref{fig8}, reveals that the field lines change their connectivity as soon as the simulation is initiated. The change in connectivity occurs because of reconnection at the X-point of HFT. 

\noindent With reconnection being known to dissipate magnetic energy, the increase of $\rm{W}^{V}_{av}$ demands additional analysis. With field line twist decreasing, the energy may increase if the net energy flux entering the sub-volume $\rm{S_{1}}$ supersedes the energy dissipation at the X-point of the HFT. Such an analysis requires estimations of Poynting flux and dissipation to high accuracy, which is presently beyond the scope of this article as stated upfront 
in the paper. Nevertheless, an attempt is made toward a coarse estimation.
A variable $|\mathbf{D}|$ is defined to approximate the $\bf{D_{B}}$ in equation (\ref{new}), indicating when and where non-ideal effects can be important. 
\begin{equation}
{|\mathbf{D}|}=\bigg{|}\derivp{\B_{l}}{t}-\nabla \times (\vec{v}_{l} \times \vec{B}_{l})\bigg{|} \label{eqD}.    
\end{equation}
Toward evaluating energy flux entering or leaving the sub-volume, an 
approximate estimation of Poynting flux is attempted with only the ideal contribution of electric field satisfying
\begin{equation}
\vec E_{\mathrm{ideal}} = - (\vec v \times \vec B) \label{r2a},   
\end{equation}
\noindent because of the ILES nature of the computation. Using straightforward vector analysis, the Poynting flux (\citealp{Kusano2002}) across the bounding surface (B) of a sub-volume can be written as
\begin{equation}
\mathrm{{S}^{B}_{av}} = \frac{1}{N}\times\displaystyle\sum_{l=0}^{N-1}{\left(|\mathbf{B}^{t}_{l}|^{2}\mathbf{v}^{n}_{l}-(\mathbf{B}^{t}_{l}\cdot\mathbf{v}^{t}_{l})\mathbf{B}^{n}_{l}\right)\cdot \Delta\mathbf{a}} \label{r2b},     
\end{equation}
where $N$ has usual meaning, $n$ and $t$ mark the normal and tangential components to the area element vector denoted by $\Delta\mathbf{a}$, respectively. Notably, $\mathrm{v_z}$ remains zero at the bottom boundary throughout the computation because of the employed boundary condition and the initial static state---see section \ref{sec41}. Consequently, only the second term contributes to the Poynting flux through the bottom boundary.

\begin{figure}[H]
\centering
\includegraphics[scale=0.15]{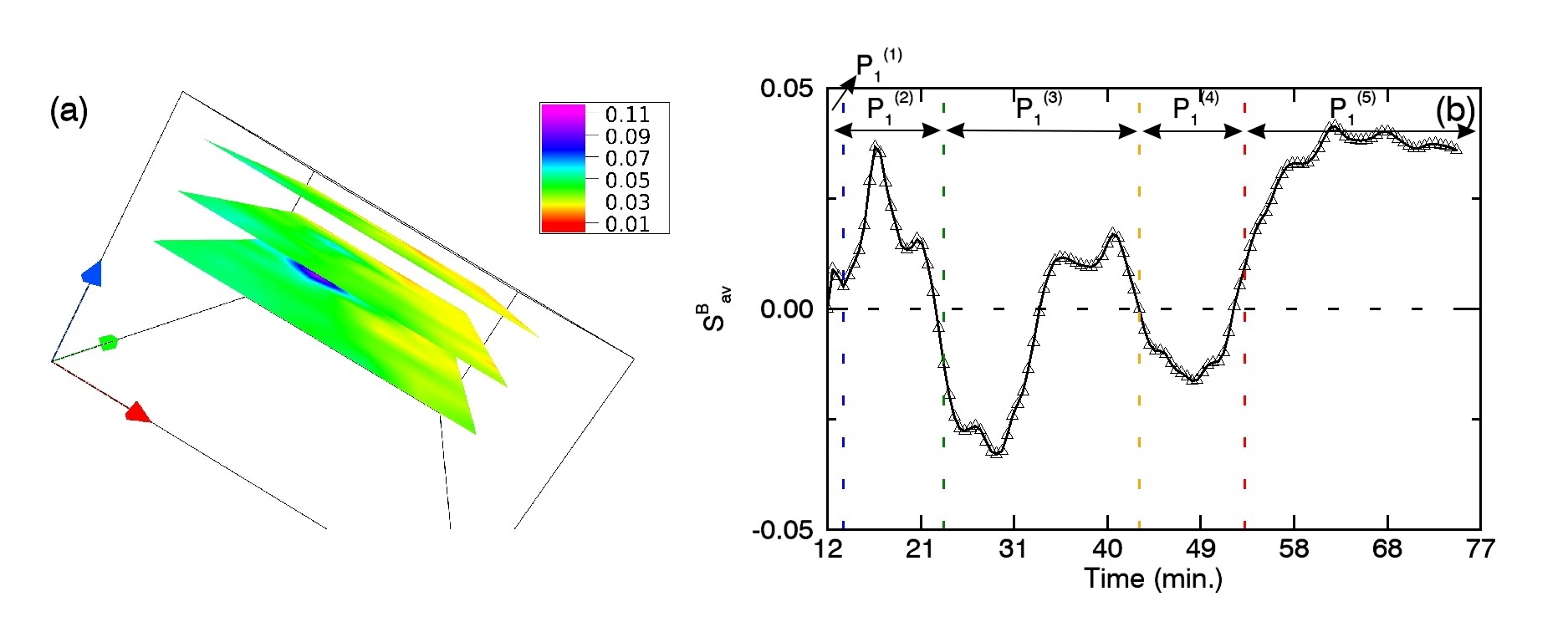}
\caption{(a) Two-dimensional data planes of temporally averaged $|{\bf{D}}|$ at three different heights in sub-volume $\rm{S_{1}}$. The mapping of data values is shown in the color bar. (b) Temporal evolution of $\mathrm{{S}^{B}_{av}}$ for sub-volume $\rm{S_{1}}$. The \textit{dashed} lines in \textit{blue}, \textit{green}, \textit{orange}, and \textit{red} colors separate the different phases. The origin of the time scale maps to 15:12 UT.}
\label{fig9}
\end{figure}

In Figure \ref{fig9}, Panel (a) plots the two-dimensional data planes of temporally averaged (averaged over the total computation time) $|{\bf{D}}|$ extracted from its 3D data volume, using the Slice Renderer function of VAPOR \citep{VAPOR2019} and $|{\bf{D}}|\in\{0.01,0.11\}$. Notably, $|{\bf{D}}|$ is largest in the neighborhood of the X-point of HFT depicted in Panel (a) of Figure \ref{fig8} and decreases away from it. The Panel (b) plots $\mathrm{{S}^{B}_{av}}$. A positive value of $\mathrm{{S}^{B}_{av}}$ indicates outflow of magnetic energy whereas negative value means energy influx.

\noindent Straightforwardly, the 
plot shows an outward energy flux up to $\rm{P_{1}^{(2)}}$, followed mostly by an inward energy flux up to $\rm{P_{1}^{(4)}}$, except for a brief time duration in $\rm{P_{1}^{(3)}}$. In the range $\rm{P_{1}^{(5)}}$, the energy flux is again outward. A comparison with magnetic energy evolution (Panel (a), Figure \ref{fig7}) shows the direction of energy fluxes to be overall consistent with energy variations for the phases $\rm{P_{1}^{(3)}}$, $\rm{P_{1}^{(4)}}$, and $\rm{P_{1}^{(5)}}$ but in complete disagreement for $\rm{P_{1}^{(2)}}$ and briefly for $\rm{P_{1}^{(3)}}$. An absolute reasoning for this disagreement is not viable within the employed framework of the model, nevertheless, a possibility is the model not being in strict adherence to the equation (\ref{eqn4}) --- leading to a violation of equation (\ref{r2a}).

\subsection{Sub-volume $\rm{S_{2}}$}
\label{sec52}

\begin{figure}[H]
\centering
\includegraphics[scale=0.15]{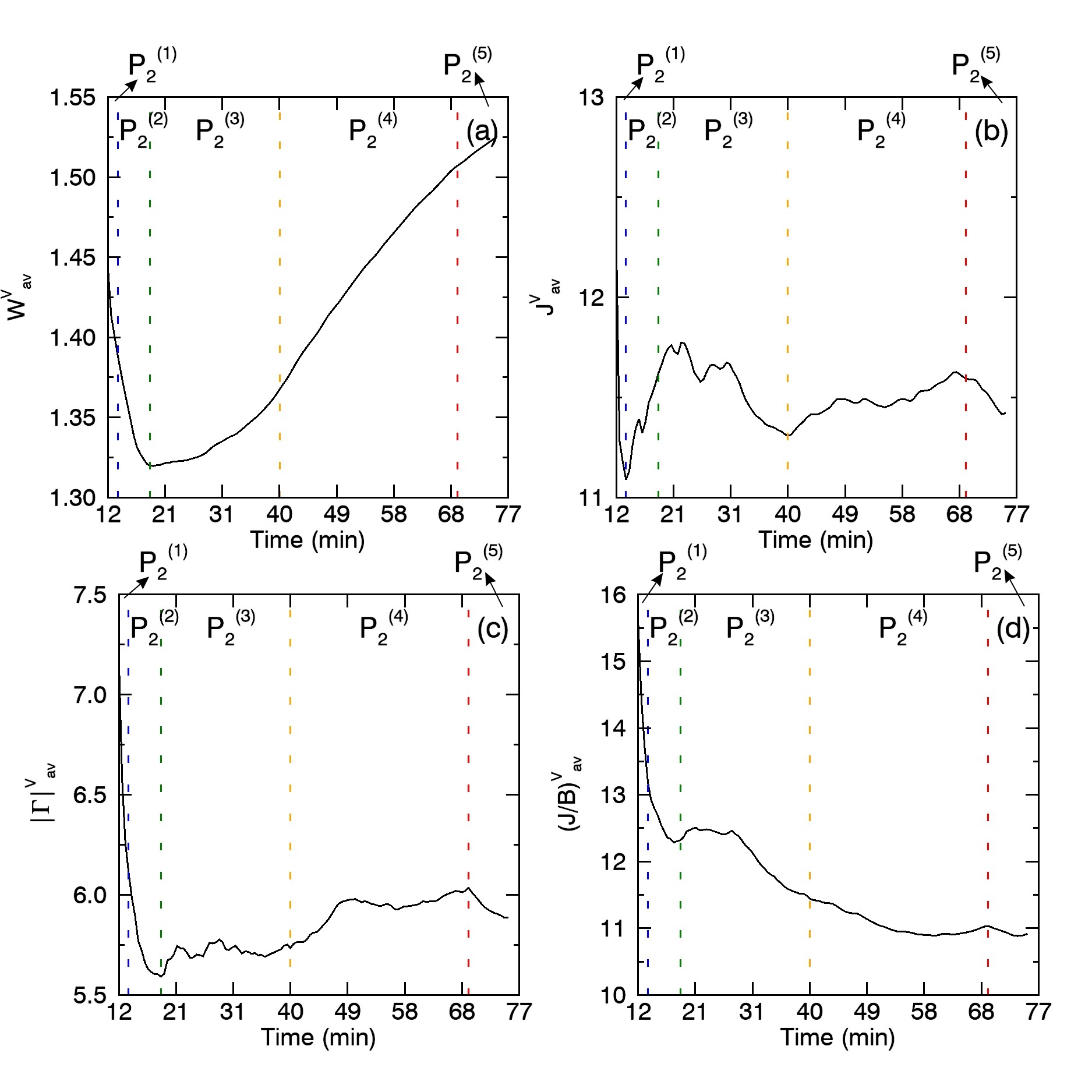}
\caption{Time evolution of grid averaged (a) magnetic energy ($\rm{W}^{V}_{av}$) (b) current density ($\rm{J}^{V}_{av}$) (c) twist parameter($\rm{|\Gamma|}^{V}_{av}$), and (d) $\rm{(J/B)}^{V}_{av}$ in sub-volume $\rm{S_{2}}$, during phases $\rm{P_{2}^{(1)}}$ (marked by the \textit{black} arrow), $\rm{P_{2}^{(2)}}$, $\rm{P_{2}^{(3)}}$, $\rm{P_{2}^{(4)}}$, and $\rm{P_{2}^{(5)}}$ (also marked by \textit{black arrow}), respectively. The \textit{dashed} lines in \textit{blue}, \textit{green}, \textit{orange}, and \textit{red} colors separate the different phases in each of the profiles. The origin of the time scale maps to 15:12 UT.}
\label{fig10}
\end{figure}

The evolution of $\rm{W}^{V}_{av}$, $\rm{J}^{V}_{av}$, $\rm{|\Gamma|}^{V}_{av}$, and $\rm{(J/B)}^{V}_{av}$ for $\rm{S}_2$ are presented in Panels (a), (b), (c), and (d) ofFigure \ref{fig10}. Again, five phases are considered, denoted by $\mathrm{P_{2}^{(i)}}$, where, $\rm{i=1,2,...,5}$. In phases $\rm{P_{2}^{(1)}}$ and $\rm{P_{2}^{(2)}}$,   $\rm{W}^{V}_{av}$, $\rm{|\Gamma|}^{V}_{av}$, and $\rm{(J/B)}^{V}_{av}$ exhibit a sharp decline. Similar behavior is observed during $\rm{P_{2}^{(1)}}$ for $\rm{J}^{V}_{av}$. Subsequently, over the span $\rm{P_{2}^{(3)}}$ to $\rm{P_{2}^{(5)}}$, there is an overall decrease in $\rm{W}^{V}_{av}$, $\rm{|\Gamma|}^{V}_{av}$ and from $\rm{P_{2}^{(2)}}$ to $\rm{P_{2}^{(5)}}$ in $\rm{J}^{V}_{av}$. However, $\rm{(J/B)}^{V}_{av}$ decreases almost continuously. The quantitative changes corresponding to different phases are summarized inFigure \ref{table3}, from which, a comparison of the terminal and initial states of the simulation reveals that the net magnetic energy in $\rm{S_{2}}$ increases, while the other parameters decrease. 

\begin{table}[h]
\caption{Summary of the quantitative changes in grid averaged profiles of magnetic energy ($\rm{W}^{V}_{av}$), current density ($\rm{J}^{V}_{av}$), twist parameter($\rm{|\Gamma|}^{V}_{av}$), and magnetic field gradient ($\rm{(J/B)}^{V}_{av}$) for sub-volume $\rm{S_{2}}$, during phases $\rm{P_{2}^{(1)}}$, $\rm{P_{2}^{(2)}}$, $\rm{P_{2}^{(3)}}$, $\rm{P_{2}^{(4)}}$, and $\rm{P_{2}^{(5)}}$, respectively. The positive and negative values indicate the rising and declining phases, while the net value in the rightmost column tells about the difference between terminal and initial states.}
\label{table3}
\begin{tabular}{ccccccc}     
\specialrule{.1em}{.05em}{.05em}
 & $\rm{P_{2}^{(1)}}$ & $\rm{P_{2}^{(2)}}$ & $\rm{P_{2}^{(3)}}$ & $\rm{P_{2}^{(4)}}$ & $\rm{P_{2}^{(5)}}$ & Net\\
\specialrule{.1em}{.05em}{.05em}
$\rm{W^{V}_{av}}$ & -0.054 & -0.069 & +0.048 & +0.140 & +0.018 & +0.083\\
$\rm{J}^{V}_{av}$ & -1.172 & +0.532 & -0.313 & +0.286 & -0.174 & -0.841\\
$\rm{|\Gamma|}^{V}_{av}$ & -1.074 & -0.521 & +0.142 & +0.300 & -0.148 & -1.301\\
$\rm{(J/B)}^{V}_{av}$ & -2.387 & -0.866 & -0.876 & -0.408 & -0.120 & -4.657\\
\specialrule{.1em}{.05em}{.05em}
\end{tabular}
\end{table}

\noindent The exploration of dynamics reveals that each layer along the vertical direction of computational box (denoted by $\rm{z_{0}=0,1,...,19}$) has nearly similar profile for magnetic energy, while for $\rm{J}^{V}_{av}$ and $\rm{|\Gamma|}^{V}_{av}$, this is not true. 

The sharp decline in $\rm{J}^{V}_{av}$ during phase $\rm{P_{2}^{(1)}}$ is predominantly caused by $\rm{z_{0}=0,1}$. The rising phase $\rm{P_{2}^{(2)}}$ is caused by the layers $\rm{z_{0}=2}$ to $9$, with dominant contributions from $\rm{z_{0}=2,3,4}$ and maximum from $\rm{z_{0}=3}$. Similarly, the declining $\rm{P_{2}^{(3)}}$ phase is shaped by layers $\rm{z_{0}=0}$ to $4$, but the most significant role is played by layers $\rm{z_{0}=1,2,3}$, while the maximum contribution arises from $\rm{z_{0}=2}$. In the later phase, i.e. $\rm{P_{2}^{(4)}}$, $\rm{J}^{V}_{av}$ increases again because of $\rm{z_{0}=11}$ to $19$. Notably, during $\rm{P_{2}^{(4)}}$, the layers $\rm{z_{0}=2}$ to $10$ display declining values of current density, thus suggesting that while current density decreases in lower layers, the overall phase is governed by the dynamical evolution in higher layers. Lastly, in the concluding phase $\rm{P_{2}^{(5)}}$, layers from $\rm{z_{0}=0}$ to $15$ exhibit decrease of current density, thus resulting in an overall decay. 

\begin{figure}[H]
\centering
\includegraphics[scale=0.18]{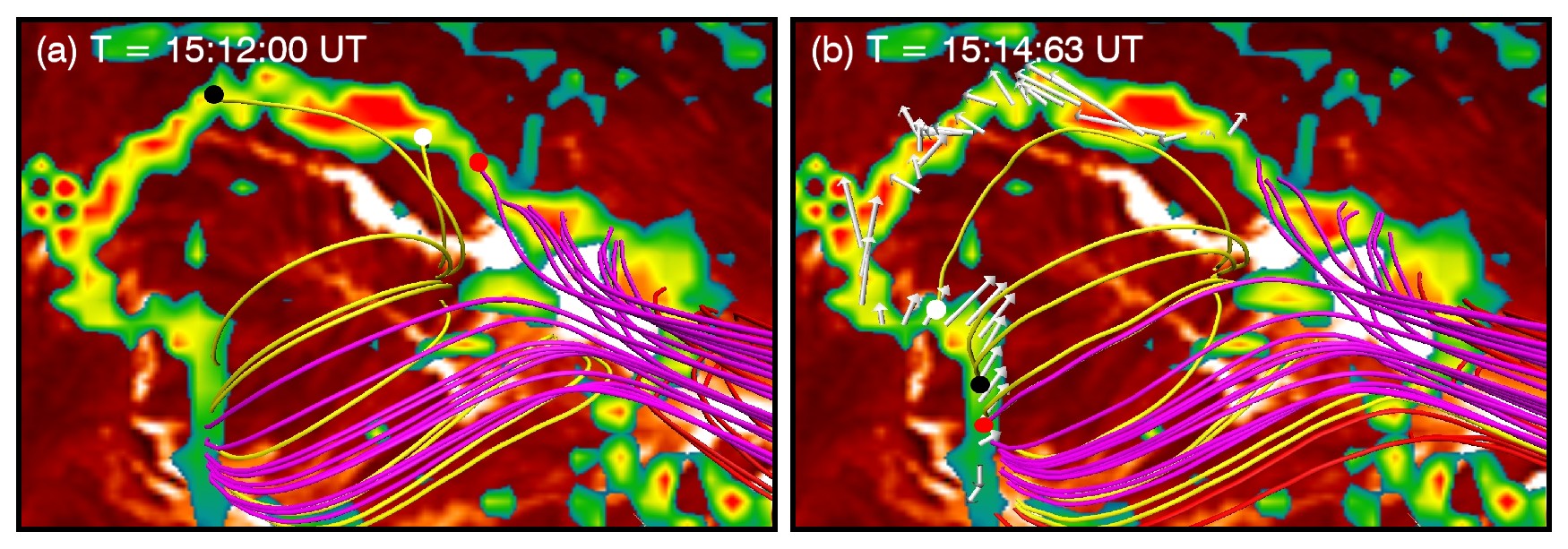}
\caption{Panels (a) and (b): Illustration of sudden shift in the footpoints of \textit{yellow} and \textit{pink} magnetic field lines within the sub-volume $\rm{S_{2}}$ due to slipping reconnection. Panel (a) highlights the initial footpoints of three selective MFLs in \textit{black}, \textit{white}, and \textit{red} colored circles. Panel (b) depicts the sudden movement of these footpoints, which is not along the direction of plasma flow (shown in white arrows). The bottom boundary is overlaid with squashing degree map and an image in 304 \AA\, channel of SDO/AIA (The spatiotemporal evolution of the magnetic field line dynamics undergoing slipping reconnection in sub-volume $\rm{S_{2}}$ is available as movie in supplementary materials).}
\label{fig11}
\end{figure}

The $\rm{|\Gamma|}^{V}_{av}$ profile reveals sharp decline during phases $\rm{P_{2}^{(1)}}$ and $\rm{P_{2}^{(2)}}$, primarily due to initial five to six layers ($\rm{z_{0}=0}$ to $5$), but as in case of $\rm{J}^{V}_{av}$, $\rm{z_{0}=0,1}$ determine the overall profile during these phases. The subsequent rising phases $\rm{P_{2}^{(3)}}$ and $\rm{P_{2}^{(4)}}$ are seen to be governed by layers $\rm{z_{0}=8}$ to $19$ and $\rm{z_{0}=12}$ to $19$, respectively. Notably, during these two phases, the lower layers, identified by $\rm{z_{0}=0}$ to $7$ and $\rm{z_{0}=0}$ to $11$ show lowering of twist over time. This behavior is reminiscent of $\rm{J}^{V}_{av}$ during $\rm{P_{2}^{(4)}}$. In the end phase $\rm{P_{2}^{(5)}}$, all except the top four layers, show lowering of twist, thus resulting in an overall decaying profile. During the early phases, i.e., up to $\rm{P_{2}^{(3)}}$ for $\rm{J}^{V}_{av}$ and $\rm{P_{2}^{(2)}}$ for $\rm{|\Gamma|}^{V}_{av}$, the lower layers ($\rm{z_{0}=0,1,...,5}$) of sub-volume $\rm{S_{2}}$ are seen to be playing the major role in determining the evolution of grid averaged parameters. This is due to the fact that the non-ideal region (the X-point of HFT) is within the first five layers of bottom boundary. Since $\rm{S_{2}}$ contains $\rm{S_{1}}$, the reconnection at X-point plays an important role during the beginning phase of  $\rm{W}^{V}_{av}$. Moreover, the energy reduction has added contribution from other sources as $\rm{S_{2}}$ covers the observed brightening $\rm{B_{2}}$ as well. We explored an instance of this possibility using the anticipated slipping reconnection in \textit{yellow} and \textit{pink} MFLs constituting the observed dome structure. Panels (a) and (b) inFigure \ref{fig11} depict a situation where sudden flipping of three selective magnetic field lines occurs (more profound in the animation provided in supplementary materials), which implies slipping reconnection. To facilitate easy identification, the footpoints of the three field lines are marked with \textit{black}, \textit{white}, and \textit{red} colored circles.

\begin{figure}[H]
\centering
\includegraphics[scale=0.18]{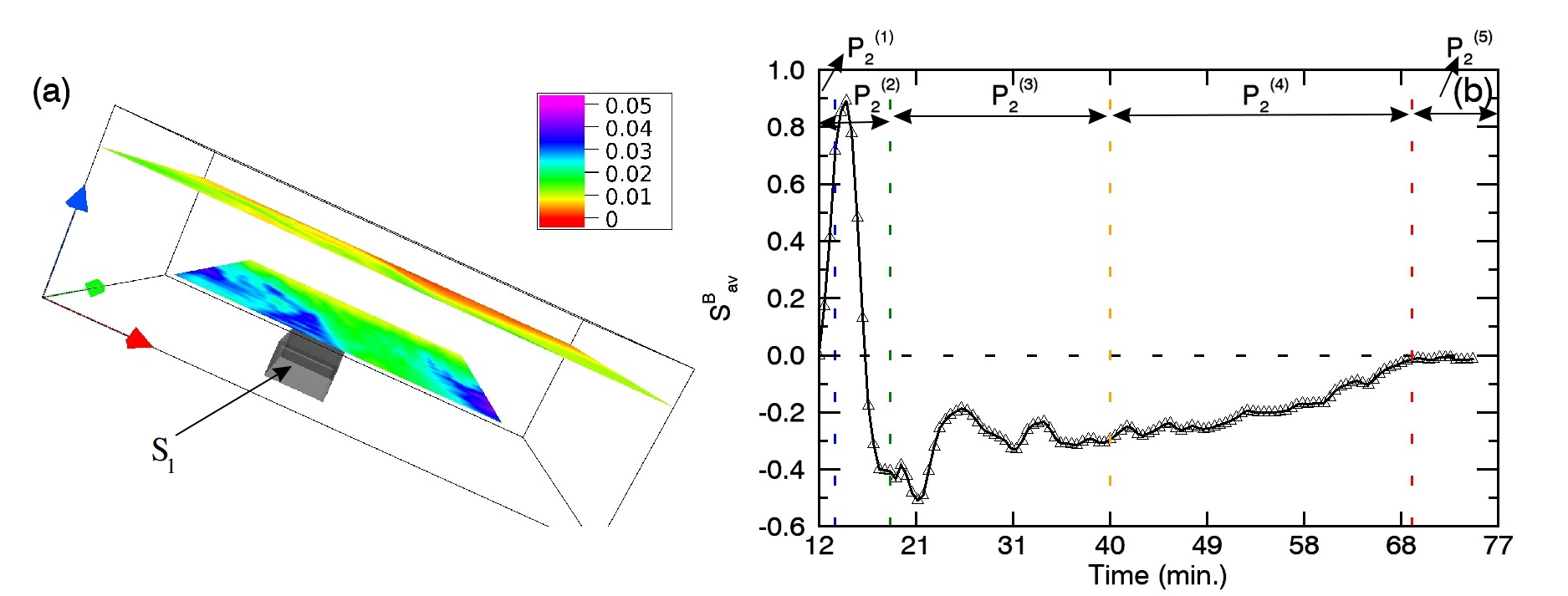}
\caption{(a) Two-dimensional data planes of temporally averaged $|{\bf{D}}|$ at two different heights in sub-volume $\rm{S_{2}}$. The mapping of data values is shown in the color bar. The \textit{black} box marks the sub-volume $\rm{S_{1}}$ (b) Temporal evolution of $\mathrm{{S}^{B}_{av}}$ for sub-volume $\rm{S_{2}}$. The \textit{dashed} lines in \textit{blue}, \textit{green}, \textit{orange}, and \textit{red} colors separate the different phases. The origin of the time scale maps to 15:12 UT.}
\label{fig12}
\end{figure}

\noindent The increase in twist from $\rm{P_{2}^{(3)}}$ to $\rm{P_{2}^{(5)}}$ is in accordance with the magnetic energy increase. To gain further insight, Figure \ref{fig12} plots the time averaged deviation $|\mathbf{D}|$ and Poynting flux in Panels (a) and (b), respectively. 

\begin{figure}[H]
\centering
\includegraphics[scale=0.18]{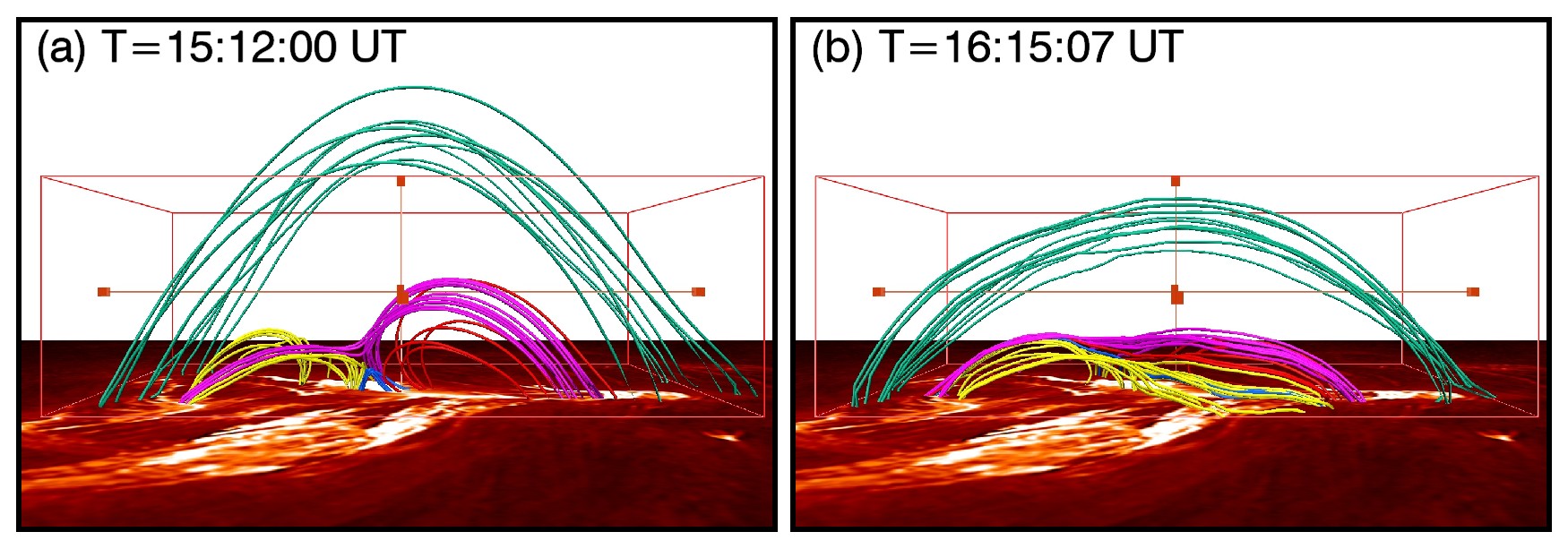}
\caption{Panels (a) and (b): Illustration of magnetic flux transfer within the sub-volume $\rm{S_{2}}$. The field lines comprising the HFT are shown, along with an additional set of \textit{green} colored MFLs, which are pushed completely inside the sub-volume $\rm{S_{2}}$ during simulation. The \textit{red} colored box marks the edges of $\rm{S_{2}}$ while the bottom boundary is overlaid with an image in 304 \AA\, channel of SDO/AIA (The spatiotemporal evolution of the magnetic field line dynamics in sub-volume $\rm{S_{2}}$ is available as movie in supplementary materials).}
\label{fig13}
\end{figure}

\noindent In $\rm{S_{2}}$ $|\mathbf{D}|\in\{0.0, 0.05\}$, which is smaller compared to that in $\rm{S_{1}}$ (marked by the black colored box in Panel (a)), signifying larger values of $|\mathbf{D}|$ to be localized at $\rm{S_{1}}$. The Poynting flux is positive for most of the $\rm{P_{2}^{(2)}}$, which is in conformity with the energy decay. For phases $\rm{P_{2}^{(3)}}$ to $\rm{P_{2}^{(5)}}$, the Poynting flux is negative, which can further be visualized from FigureFigure \ref{fig13}, where a portion of \textit{green} field lines are pushed completely inside $\rm{S_{2}}$ (\textit{red} colored box). The corresponding energy influx along with the increment in twist seems to overwhelm dissipation, thus resulting in the observed energy increase. 

\subsection{Sub-volume $\rm{S_{3}}$}
\label{sec51}

\begin{figure}[H]
\centering
\includegraphics[scale=0.15]{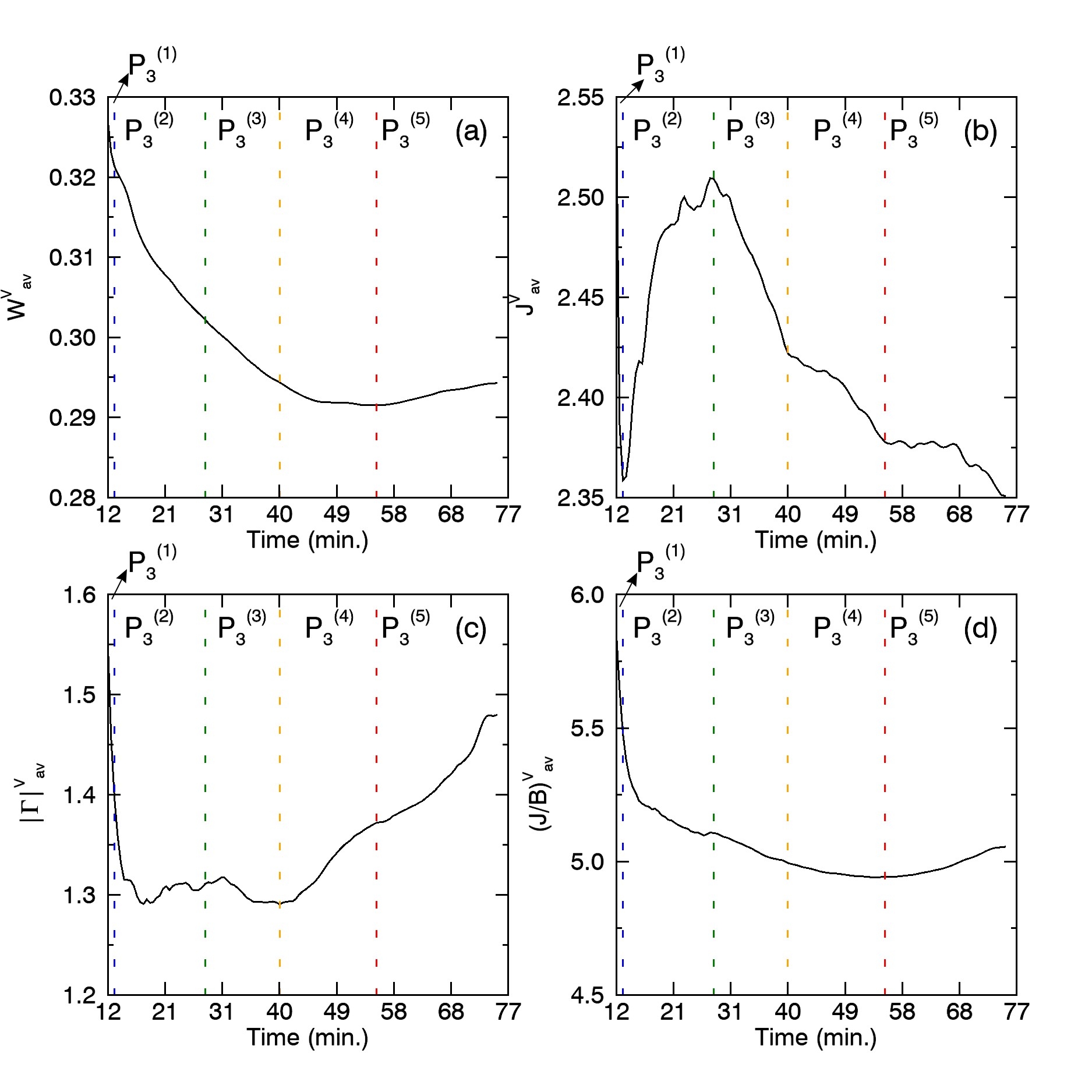}
\caption{Time evolution of grid averaged (a) magnetic energy ($\rm{W}^{V}_{av}$) (b) current density ($\rm{J}^{V}_{av}$) (c) twist parameter($\rm{|\Gamma|}^{V}_{av}$), and (d) $\rm{(J/B)}^{V}_{av}$ in sub-volume $\rm{S_{3}}$ during phases $\rm{P_{3}^{(1)}}$ (marked by the \textit{black} arrow), $\rm{P_{3}^{(2)}}$, $\rm{P_{3}^{(3)}}$, $\rm{P_{3}^{(4)}}$, and $\rm{P_{3}^{(5)}}$, respectively. The \textit{dashed} lines in \textit{blue}, \textit{green}, \textit{orange}, and \textit{red} colors separate the different phases in each of the profiles. The origin of the time scale maps to 15:12 UT.}
\label{fig14}
\end{figure}

The sub-volume $\rm{S_{3}}$ encompasses the complete extent of the observed brightening. For convenience, the evolution in $\rm{S_{3}}$ is investigated in five phases, defined by $\mathrm{P_{3}^{(i)}}$, where, $\rm{i=1,2,...,5}$. The temporal evolution of the grid averaged parameters is shown inFigure \ref{fig14}. 

\noindent Panel (a) reveals that $\rm{S_{3}}$ exhibits continuous decrease in $\rm{W}^{V}_{av}$ up to $\rm{P_{3}^{(4)}}$, which is in close agreement with the end time of the flare. Such uninterrupted decrement is a prime signature of relaxation in the considered volume. From Panel (b), it is seen that after an initial drop, $\rm{J}^{V}_{av}$ peaks at 15:27 UT, subsequently followed by a declining profile.  Panel (c) indicates that $\rm{|\Gamma|}^{V}_{av}$ decays up to 15:39 UT, which nearly corresponds to the peak time of the flare. This suggests lowering of overall twist and hence a simplification of field line complexity, which further complements the interpretation of relaxation within the sub-volume. In the later phase, there is an increase in twist while the magnetic field gradient (Panel (d)) is seen to be declining continuously with very small increment toward the end of simulation. Overall, the volume averaged MHD evolution in $\rm{S_{3}}$ is similar to the overall simulated dynamics. The quantitative changes associated with the grid averaged profiles are summarized inFigure \ref{table4}. Notably, in this sub-volume, the terminal state is characterized by a reduced value of all the parameters, i.e. magnetic energy, current density, twist, and magnetic field gradients.\\

\begin{table}[h]
\caption{Summary of the quantitative changes in grid averaged profiles of magnetic energy ($\rm{W}^{V}_{av}$), current density ($\rm{J}^{V}_{av}$), twist parameter($\rm{|\Gamma|}^{V}_{av}$), and magnetic field gradient ($\rm{(J/B)}^{V}_{av}$) for sub-volume $\rm{S_{3}}$, during phases $\rm{P_{3}^{(1)}}$, $\rm{P_{3}^{(2)}}$, $\rm{P_{3}^{(3)}}$, $\rm{P_{3}^{(4)}}$, and $\rm{P_{3}^{(5)}}$, respectively. The positive and negative values indicate the rising and declining phases, while the net value in the rightmost column tells about the difference between terminal and initial states.}
\label{table4}
\begin{tabular}{ccccccc}     
\specialrule{.1em}{.05em}{.05em}
 & $\rm{P_{3}^{(1)}}$ & $\rm{P_{3}^{(2)}}$ & $\rm{P_{3}^{(3)}}$ & $\rm{P_{3}^{(4)}}$ & $\rm{P_{3}^{(5)}}$ & Net\\
\specialrule{.1em}{.05em}{.05em}
$\rm{W^{V}_{av}}$ & -0.005 & -0.019 & -0.008 & -0.003 & +0.003 & -0.033\\
$\rm{J}^{V}_{av}$ & -0.185 & +0.151 & -0.087 & -0.044 & -0.027 & -0.192\\
$\rm{|\Gamma|}^{V}_{av}$ & -0.151 & -0.086 & -0.020 & +0.081 & +0.107 & -0.069\\
$\rm{(J/B)}^{V}_{av}$ & -0.373 & -0.374 & -0.112 & -0.052 & +0.113 & -0.800\\
\specialrule{.1em}{.05em}{.05em}
\end{tabular}
\end{table}

\begin{figure}[H]
\centering
\includegraphics[scale=0.18]{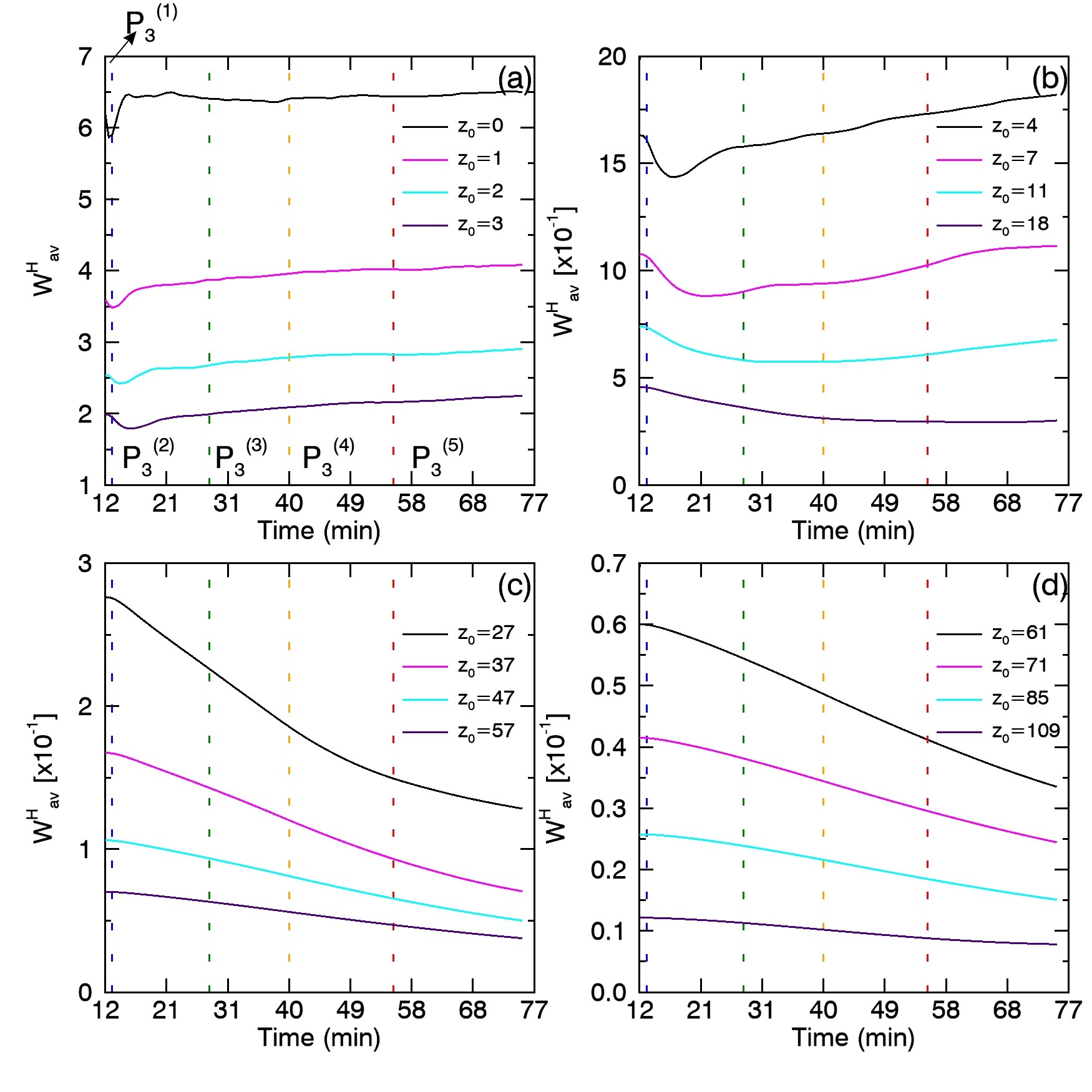}
\caption{Time evolution of $\rm{W}^{H}_{av}$ for sub-volume $\rm{S_{3}}$ at different $\rm{z=z_{0}}$ layers, shown by the \textit{black}, \textit{magenta}, \textit{cyan}, and \textit{indigo} color \textit{solid} lines. The labels indicate the chosen $\rm{z_{0}}$ value in each Panel. The different phases $\mathrm{P_{3}^{(i)}}$, where $\rm{i=0,1,...,5}$, are marked only in Panel (a) to avoid clutter, while the \textit{dashed} lines separating the phases are marked in each of the Panels. The \textit{y}-scale in Panels (b), (c), and (d) differs by an order of magnitude ($10^{-1}$) than in Panel (a). The origin of time on x-axis maps to 15:12 UT.}
\label{fig15}
\end{figure}

\noindent Notably, sub-volume $\rm{S_{3}}$ spans the full extent of observed brightenings and the full vertical extent of the computational box. Consequently, to understand the simulated dynamics more comprehensively, an analysis of horizontally averaged magnetic energy ($\rm{W}^{H}_{av}$), current density ($\rm{J}^{H}_{av}$), and twist parameter ($\rm{|\Gamma|}^{H}_{av}$) is carried out. These follow the definitions given in \ref{nd1}, \ref{nd2}, and \ref{nd4} but grid averaged over different $\rm{z=z_{0}}$ layers, each having $N=70\times 60$ voxels along the $x-$ and $y-$directions, respectively.Figure \ref{fig15} shows the temporal profile of $\rm{W}^{H}_{av}$ for selected layers. During phase $\rm{P_{3}^{(1)}}$, all the layers exhibit decreasing $\rm{W}^{H}_{av}$ with significant contribution from $\rm{z_{0}=0,1,2}$, thus leading to declining $\rm{W}^{V}_{av}$. Further, as shown in Panels (a) and (b), the subsequent phases are seen to have increasing $\rm{W}^{H}_{av}$ for $\rm{z_{0}=0}$ to $3$ in $\rm{P_{3}^{(2)}}$, $\rm{z_{0}=1}$ to $10$ in $\rm{P_{3}^{(3)}}$, $\rm{z_{0}=0}$ to $14$ in $\rm{P_{3}^{(4)}}$, and $\rm{z_{0}=0}$ to $17$ in $\rm{P_{3}^{(5)}}$, respectively. Note that the evolution of $\rm{W}^{H}_{av}$ differs by an order of magnitude in the two Panels. The remaining layers during these phases exhibit declining $\rm{W}^{H}_{av}$, as evident from Panels (b), (c), and (d). In effect then, owing to their larger number, these remaining layers dominate the profile evolution of $\rm{W}^{V}_{av}$ during phases $\rm{P_{3}^{(2)}}$, $\rm{P_{3}^{(3)}}$, and $\rm{P_{3}^{(4)}}$. However, in the end phase, the dynamics in $\rm{z_{0}=0}$ to $17$ takes control, thus leading to increasing $\rm{W}^{V}_{av}$ during $\rm{P_{3}^{(5)}}$.

\begin{figure}[H]
\centering
\includegraphics[scale=0.18]{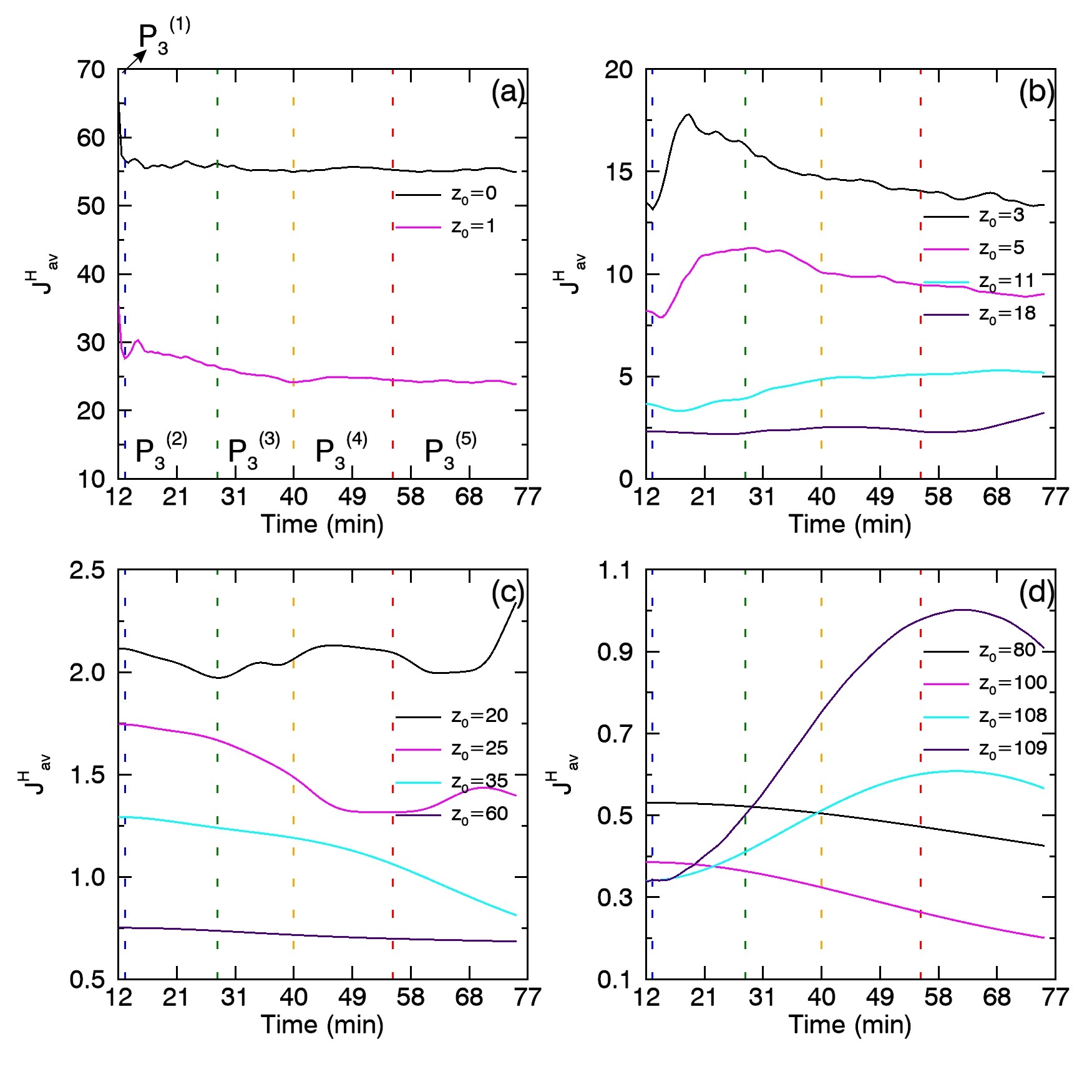}
\caption{Time evolution of $\rm{J}^{H}_{av}$ for sub-volume $\rm{S_{3}}$ at different $\rm{z=z_{0}}$ layers, shown by the \textit{black}, \textit{magenta}, \textit{cyan}, and \textit{indigo} color \textit{solid} lines. The labels indicate the chosen $\rm{z_{0}}$ value in each Panel. The different phases $\mathrm{P_{3}^{(i)}}$, where $\rm{i=0,1,...,5}$, are marked only in Panel (a) to avoid clutter, while the \textit{dashed} lines separating the phases are marked in each of the Panels. The origin of time on x-axis maps to 15:12 UT.}
\label{fig16}
\end{figure}

\noindent Due to larger volume of $\rm{S_{3}}$, the resulting $\rm{W}^{V}_{av}$ profile is jointly governed by both larger (smaller) decrements in the lower (higher) layers. However, for $\rm{S_{2}}$, whose vertical extent is restricted to $\rm{z_{0}=19}$, layers from Panel (a) and partly from Panel (b) can be visualized to jointly reproduce an initial fall, followed by continuous rise. Next, the behavior of $\rm{J}^{H}_{av}$ is explored, as shown inFigure \ref{fig16}.  

\begin{figure}[H]
\centering
\includegraphics[scale=0.18]{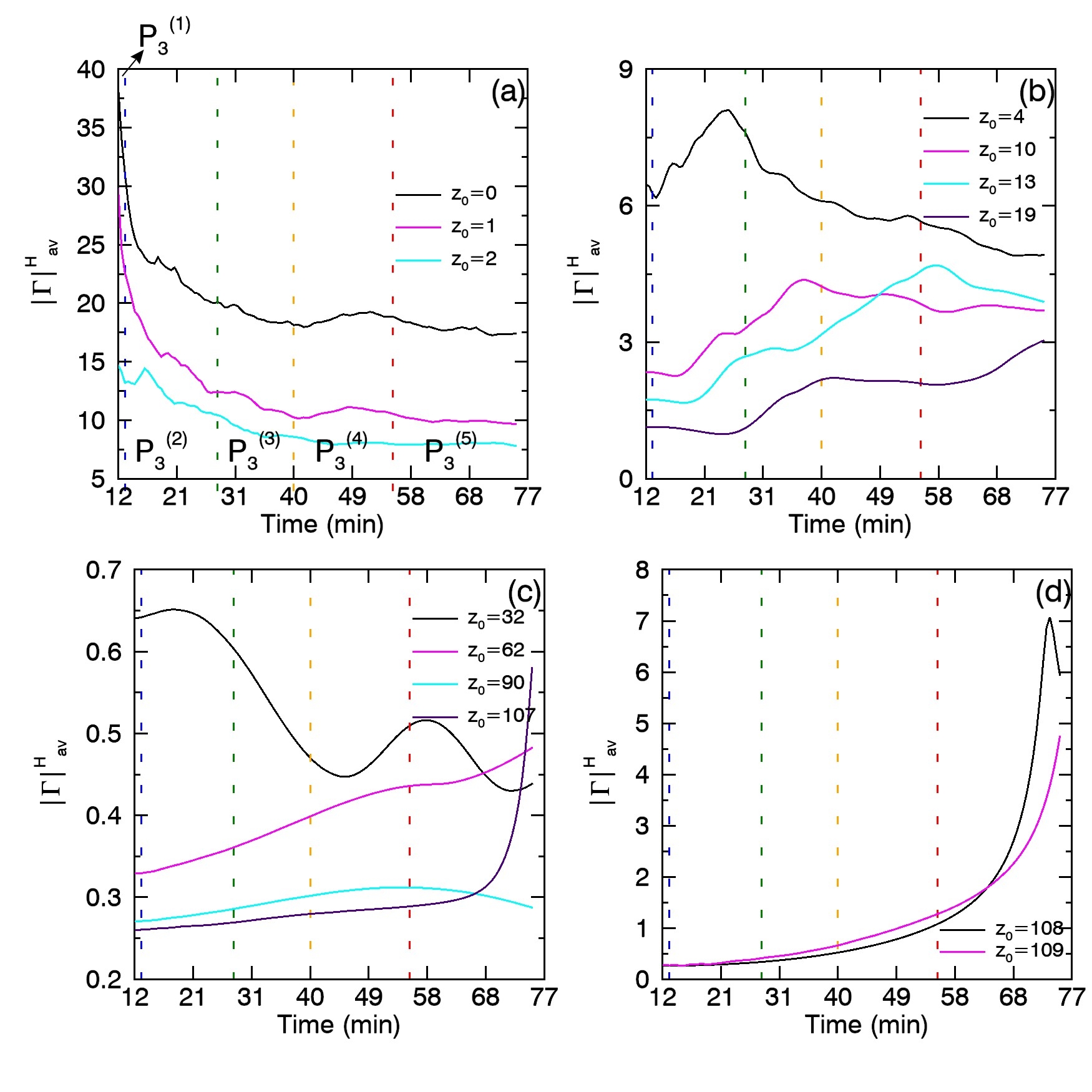}
\caption{Time evolution of $\rm{|\Gamma|}^{H}_{av}$ for sub-volume $\rm{S_{3}}$ at different $\rm{z=z_{0}}$ layers, shown by the \textit{black}, \textit{magenta}, \textit{cyan}, and \textit{indigo} color \textit{solid} lines. The labels indicate the chosen $\rm{z_{0}}$ value in each Panel. The different phases $\mathrm{P_{3}^{(i)}}$, where $\rm{i=0,1,...,5}$, are marked only in Panel (a) to avoid clutter, while the \textit{dashed} lines separating the phases are marked in each of the Panels. The origin of time on x-axis maps to 15:12 UT.}
\label{fig17}
\end{figure}

\noindent During $\rm{P_{3}^{(1)}}$, other than the top two, all layers exhibit decreasing $\rm{J}^{H}_{av}$, thereby causing the sharp decline of $\rm{J}^{V}_{av}$ in this phase. Notably, the dominant role is played by the bottom layers $\rm{z_{0}=0,1}$, as may be seen from Panel (a). Subsequently, in the next two phases, the process of current formation and dissipation within the HFT governs the evolution. As depicted in Panels (a) and (b), the increase in $\rm{J}^{V}_{av}$ during $\rm{P_{3}^{(2)}}$ is essentially due rising $\rm{J}^{H}_{av}$ in layers $\rm{z_{0}=2}$ to $11$. Similarly, the decreasing $\rm{J}^{H}_{av}$ in $\rm{z_{0}=0}$ to $7$ causes the decline of $\rm{J}^{V}_{av}$ during phase $\rm{P_{3}^{(3)}}$ despite the increasing $\rm{J}^{H}_{av}$ in $\rm{z_{0}=8}$ to $20$. In the remaining two phases $\rm{P_{3}^{(4)}}$ and $\rm{P_{3}^{(5)}}$, the segregation of any dominant contribution from $\rm{z=z_{0}}$ layers was found to be difficult. However, the $\rm{J}^{V}_{av}$ profile is understood from the finding that $\rm{J}^{H}_{av}$ decreases significantly in layers $\rm{z_{0}=23}$ to $107$ and $\rm{z_{0}=26}$ to $109$, respectively, as evident from Panels (c) and (d). Interestingly, the two topmost layers reveal an abrupt increase in $\rm{J}^{H}_{av}$, an understanding of which requires investigation of field line dynamics. Lastly, the behavior of $\rm{|\Gamma|}^{H}_{av}$ is investigated, as depicted inFigure \ref{fig17}.

\begin{figure}[H]
\centering
\includegraphics[scale=0.5]{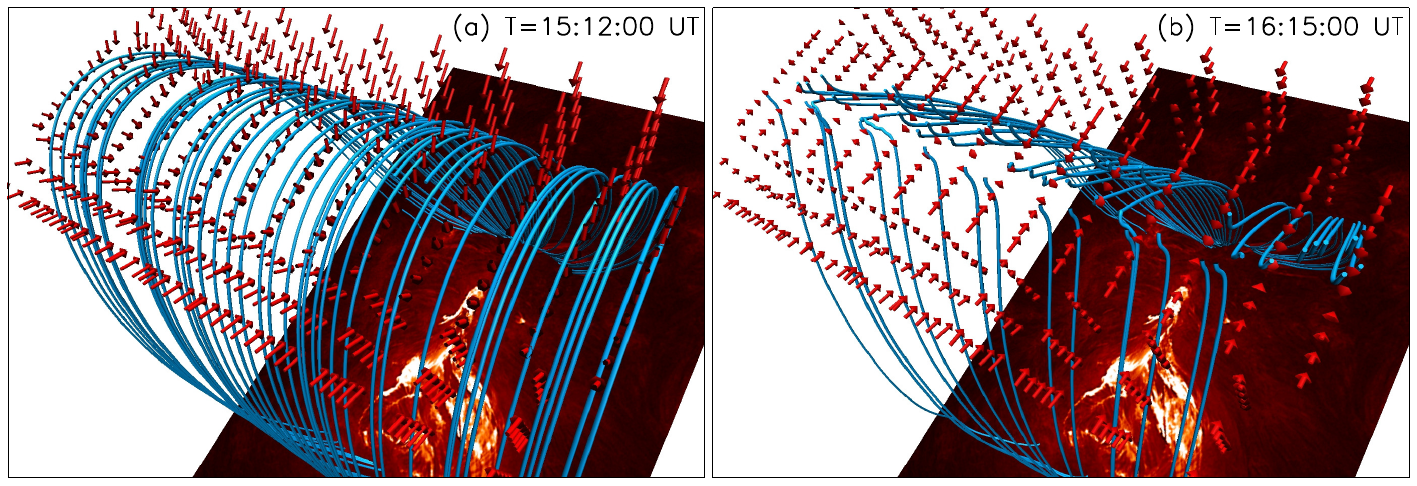}
\caption{Illustration of magnetic field line dynamics responsible for abrupt rise of $\rm{J}^{H}_{av}$ and $\rm{|\Gamma|}^{H}_{av}$ in the top two layers of the computational box. The \textit{blue} MFLs and \textit{red} arrows in Panel (a) depict the bipolar potential field lines and direction of Lorentz force in the beginning of simulation. Panel (b) depicts the deformation in MFLs due to action of Lorentz force over the course of simulation. The bottom boundary is overlaid with image in 304 \AA\, channel of SDO/AIA (The spatiotemporal evolution of these \textit{blue} MFLs is available as movie in supplementary material).}
\label{fig18}
\end{figure}

\noindent Panels (a) and (b) reveal that $\rm{|\Gamma|}^{H}_{av}$ decreases for $\rm{z_{0}=0}$ to $11$ and increases for $\rm{z_{0}=3}$ to $16$ during phases $\rm{P_{3}^{(1)}}$ and $\rm{P_{3}^{(2)}}$, respectively. However, $\rm{|\Gamma|}^{V}_{av}$ declines during both the phases due to the dominating decrease of $\rm{|\Gamma|}^{H}_{av}$ in the bottom layers, i.e. $\rm{z_{0}=0,1,2}$. A similar behavior is observed for phase $\rm{P_{3}^{(3)}}$, where the fall of $\rm{|\Gamma|}^{H}_{av}$ in $\rm{z_{0}=0}$ to $7$ dominates the rise of $\rm{|\Gamma|}^{H}_{av}$ in $\rm{z_{0}=8}$ to $23$. The increase of $\rm{|\Gamma|}^{V}_{av}$ during $\rm{P_{3}^{(4)}}$ is seen to be consequence of dynamics in $\rm{z_{0}=11}$ to $16$ (Panel (b)) and $\rm{z_{0}=20}$ to $32$ (Panel (c)). In the concluding phase $\rm{P_{3}^{(5)}}$, $\rm{|\Gamma|}^{V}_{av}$ increases further owing to increasing $\rm{|\Gamma|}^{H}_{av}$ in $\rm{z_{0}=14}$ to $20$ (Panel (b)) and the abrupt increase of $\rm{|\Gamma|}^{H}_{av}$ in the topmost layers, namely $\rm{z_{0}=107,108,109}$ (Panel (d)). The underlying reason for this abrupt rise may be understood fromFigure \ref{fig18}. Panel (a) depicts \textit{blue} MFLs which constitute the potential bipolar loops, while the \textit{red} arrows show direction of Lorentz force around the topmost region of computational box. As evident from Panel (b), the converging force pushes magnetic field lines toward each other, which leads to stressing of the configuration. Notably, it is not clear whether the disconnected MFLs in Panel (b) are a consequence of reconnection or movement of field lines outward from the box. Such a discontinuity results in a large gradient and hence sudden rise in $\rm{J}^{H}_{av}$ and $\rm{|\Gamma|}^{H}_{av}$.  

\begin{figure}[H]
\centering
\includegraphics[scale=0.17]{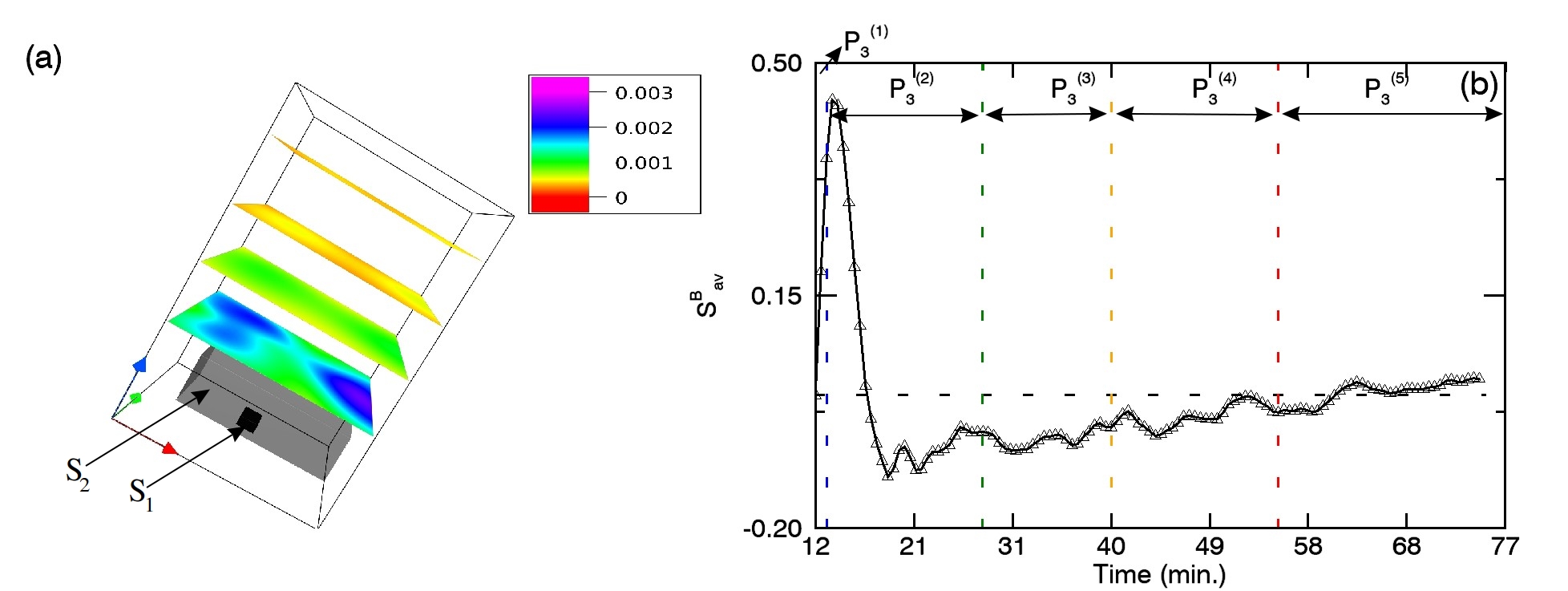}
\caption{(a) Two-dimensional data planes of temporally averaged $|{\bf{D}}|$ at different heights in sub-volume $\rm{S_{3}}$. The mapping of data values is shown in the color bar. The \textit{black} boxes mark the sub-volumes $\rm{S_{1}}$ and $\rm{S_{2}}$ (b) Temporal evolution of $\mathrm{{S}^{B}_{av}}$ for sub-volume $\rm{S_{3}}$. The \textit{dashed} lines in \textit{blue}, \textit{green}, \textit{orange}, and \textit{red} colors separate the different phases. The origin of the time scale maps to 15:12 UT.}
\label{fig19}
\end{figure}

\noindent Figure \ref{fig19} plots slice rendering of time averaged $|{\bf{D}}|$ along with the Poynting flux. The $|{\bf{D}}|\in\{0,0.003\}$, which is one and two orders less than its values in $\rm{S_{2}}$ and $\rm{S_{1}}$ (marked in Panel (a) with arrows), respectively. Comparison of $|{\bf{D}}|$ in all the three sub-volumes indicates localization of maximal $|{\bf{D}}|$ at $\rm{S_{1}}$ and specifically, at the neighborhood of the X-point---the primary reconnection site. Such localization of $|{\bf{D}}|$ is compatible with the general idea of ILES. On an average the Poynting flux is $\approx 30\%$ of its value for 
$\mathrm{S_{2}}$ and is predominantly negative, implying energy influx. As a consequence, the decrease in magnetic energy can be attributed to the overall decrease in twist conjointly with non-ideal effects, contributed primarily from $\rm{S_{1}}$ and further augmented by slipping reconnections in $\rm{S_{2}}$.

\subsection{Extent of Magnetic Relaxation}
\label{TTP}

\begin{figure}[H]
\centering
\includegraphics[scale=0.12]{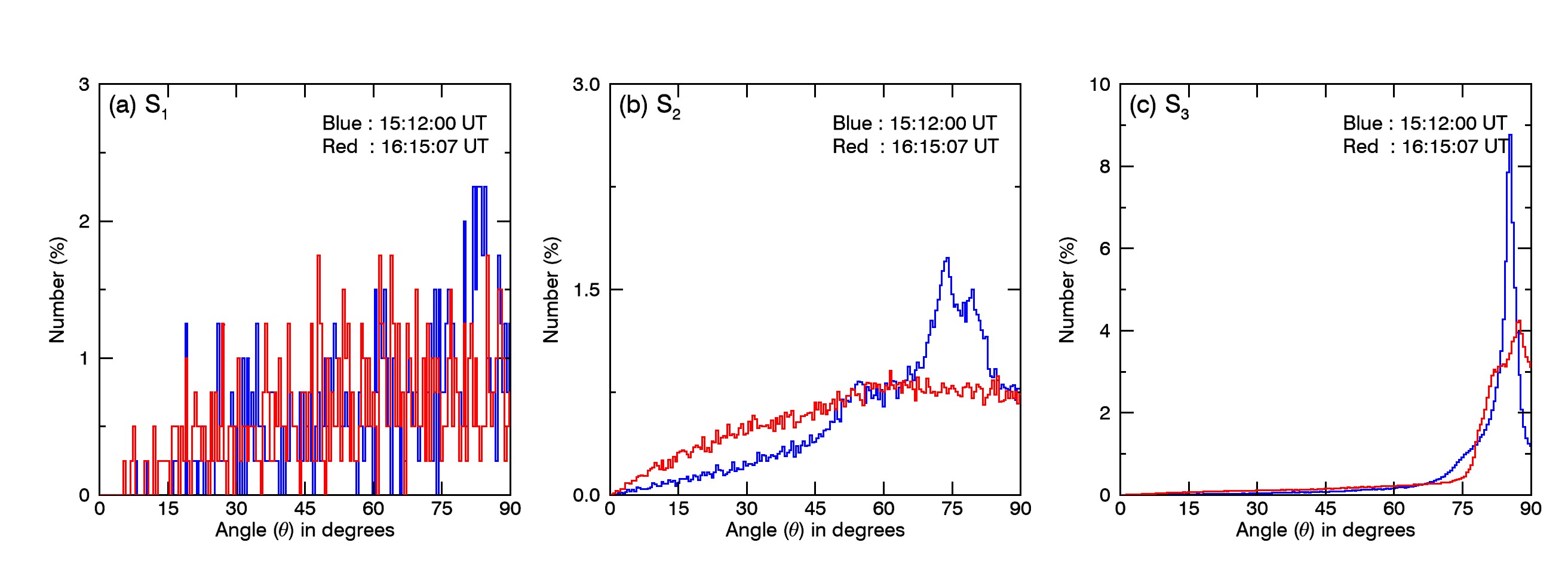}
\caption{Distribution of angles between current density ($\mathbf{J}$) and magnetic field vectors ($\mathbf{B}$) in sub-volumes $\rm{S_{1}}$, $\rm{S_{2}}$, and $\rm{S_{3}}$ at the beginning (\textit{blue}) and end (\textit{red}) of simulation.}
\label{fig20}
\end{figure}

As describe earlier, force-free states are characterized by field aligned current density. Further, the distribution of $\alpha$ distinguishes between the nonlinear and linear force-free states. Consequently, in order to check the extent of relaxation in our simulation, we compare the histogram of angle ($\rm{\theta}$) between $\mathbf{J}$ and $\mathbf{B}$ at the beginning and end of simulation for each of the sub-volumes. The $\theta$ plots inFigure \ref{fig20} utilize the transformation $180^{\circ}-\rm{\theta}$ to map $\rm{\theta} \geq 90^{\circ}$ in the range $0^{\circ} \leq \rm{\theta} \leq 90^{\circ}$. Panels (a) and (b) for sub-volumes $\rm{S_{1}}$ and $\rm{S_{2}}$ reveal wide distributions extending over the entire range of angles for both the time instants. On the other hand, Panel (c) for $\rm{S_{3}}$ shows comparatively narrow distributions peaking around $90^{\circ}$. Presumably, this is due to the fact that $\rm{S_{3}}$ spans the full vertical extent of the computational box and the variation of $\rm{\theta}$ along height in non-force-free extrapolation model exhibits an increasing trend up to $90^{\circ}$ (seeFigure \ref{fig3}). Due to small size of $\rm{S_{1}}$ and hence limited number of voxels, we could not identify any trend in the variation of $\rm{\theta}$ with time except that the distribution is wide, which does not support the presence of field aligned current in terminal state. However, careful comparison of the \textit{blue} and \textit{red} profiles in Panels (b) and (c) suggests that during simulation, fraction of voxels with $\rm{\theta}\geq 60^{\circ}$ in $\rm{S_{2}}$ and $\rm{S_{3}}$ decrease, which we estimate to be 20\% and 24\%, respectively. This suggests that the magnetic configuration tends to relax towards a force-free state. However, in the present simulation, neither the wide distribution in $\rm{S_{2}}$ nor the narrow distribution centered around $\rm{\theta}=90^{\circ}$ in $\rm{S_{3}}$ support a strictly field aligned current density. 

\begin{figure}[H]
\centering
\includegraphics[scale=0.18]{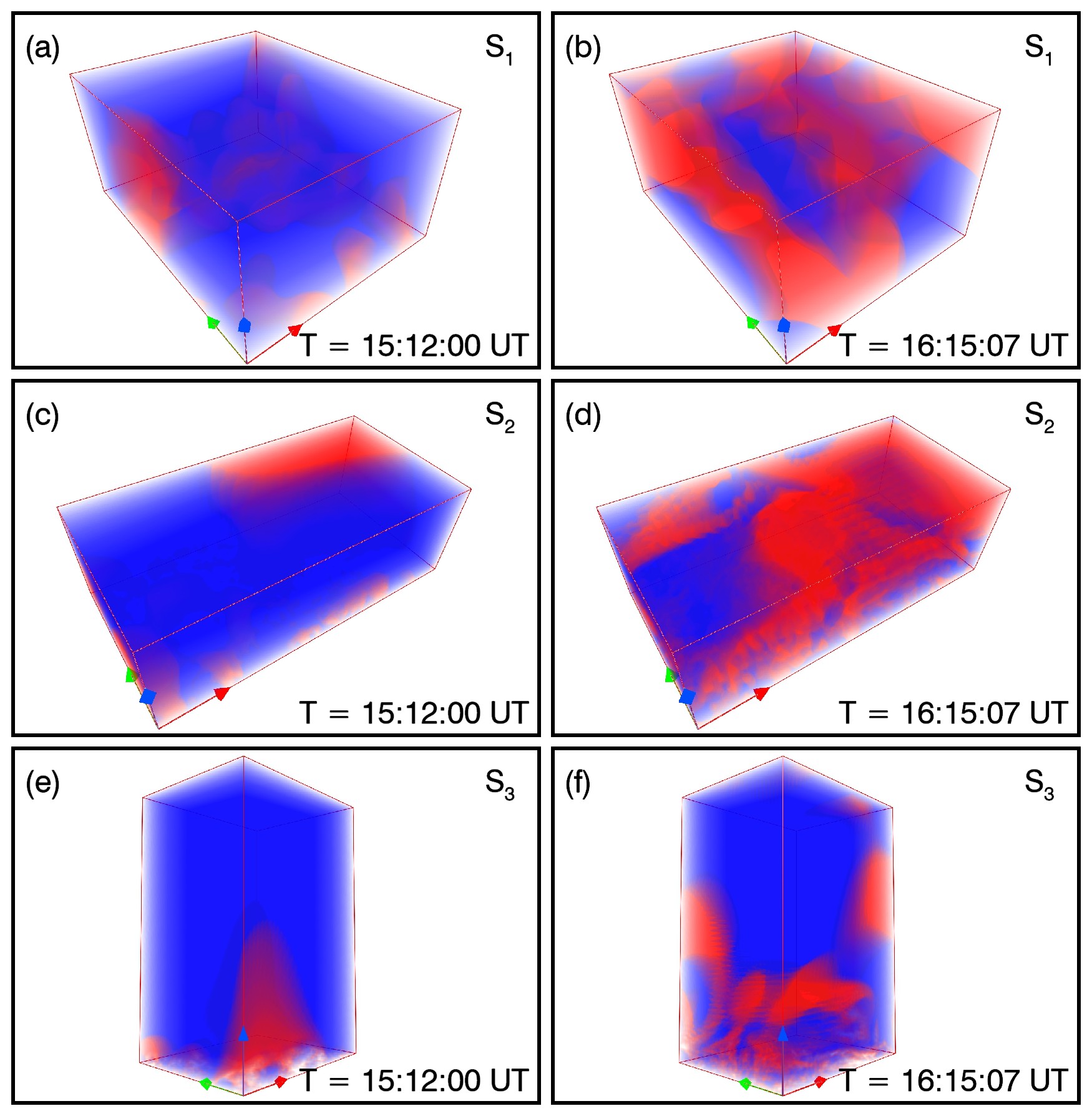}
\caption{Direct Volume Rendering (DVR) of the twist parameter ($\Gamma$) for sub-volumes $\rm{S_{1}}$, $\rm{S_{2}}$, and $\rm{S_{3}}$ at the initial and terminal state of numerical simulation. The \textit{blue} and \textit{red} colors represent positive and negative values of $\Gamma$ (The spatiotemporal evolution of distribution of twist in each sub-volume is available as a movie in supplementary materials).}
\label{fig21}
\end{figure}

\noindent Therefore, in our case, the terminal state of the simulation remains in non-equilibrium, suggesting that further magnetic relaxation is possible. To check this, we carried out an auxiliary simulation, extending its time duration to twice that of the original one. When integrated over the whole computational domain, the grid averaged angle drops by $5.7^{\circ}$ ($64.32^{\circ}$ to $58.62^{\circ}$) in the auxiliary simulation, as compared to $4.3^{\circ}$ ($64.32^{\circ}$ to $60.01^{\circ}$) in the original simulation, validating the possibility of further relaxation. Furthermore, we explored the time evolution of the twist parameter ($\Gamma$) for each of the sub-volumes, as shown inFigure \ref{fig21}. The \textit{red} and \textit{blue} colors represent the negative and positive values, respectively. We note that at the initial time instant, the distribution is dominated by positive $\Gamma$ for each sub-volume, as shown in Panels (a),(c), and (e). As the simulation progresses, negative $\Gamma$ begins to increase and terminal state consists of both positively and negatively signed values. The noteworthy aspect is the progressively increasing intermixing of the \textit{blue} and \textit{red} colors, exhibiting gradual fragmentation (better visualized in the movie provided with supplementary materials) into smaller structures, as evident from Panels (d) and (f). Such fragmentation is indicative of development of turbulence (e.g. \citealp{2011P}, also see \citealp{Veltri2009}) but since, a quantitative investigation regarding the extent of developed turbulence is presently beyond the scope of this work, it is difficult to comment on this aspect further. 

\section{Summary and Discussion}
\label{sec6}

In this study, we explore the process of magnetic relaxation during an observed solar flare. For the purpose, an extrapolated magnetic field is utilized as input to carry out a data-based numerical simulation. We select a M1.3 class flare on 04 January, 2015, hosted by the active region NOAA 12253, spanning a time of nearly 35 minutes. Observations of the flare in 1600 \AA\, and 304 \AA\, channels of SDO/AIA (seeFigure \ref{fig2}) reveal the spatiotemporal evolution of identified brightenings (namely $\rm{B_{1}}$ and $\rm{B_{2}}$), the existence of chromospheric flare ribbons, and a dome shaped structure toward the eastern side. To explore the potential reconnection sites and their morphologies, the vector magnetogram at 15:12 UT from SDO/HMI is employed to extrapolate the magnetic field using a non-force-free field (NFFF) model. We find a hyperbolic flux tube (HFT), overlying the dome structure and the brightenings $\rm{B_{1}}$, $\rm{B_{2}}$. In general, identification of all the individual reconnection sites within the computational volume is a non-trivial exercise and since, the HFT is spatially correlated with the observed brightenings, we envisage it as the primary reconnection site and focus on it for further analysis. 
To investigate reconnection dynamics and magnetic relaxation, the EULAG-MHD model is employed to execute an Implicit Large Eddy simulation (ILES), which uses the extrapolated magnetic field as initial condition. The simulation relies on the intermittent and localized dissipative property of the advection scheme MPDATA which smooths out under-resolved variables and numerically replicates magnetic reconnection. Although quantification of this dissipation is strictly meaningful only in the spectral space, nevertheless a rough estimation is carried out by analyzing $|{\bf{D}}|$: the deviation of induction equation from its ideal form.  

Toward exploring magnetic relaxation, we consider the temporal evolution of grid averaged parameters such as magnetic energy ($\mathrm{{W}^{V}_{av}}$), twist parameter ($\mathrm{|\Gamma|^{V}_{av}}$), magnetic field gradient ($\mathrm{{(J/B)}^{V}_{av}}$), and current density ($\mathrm{{J}^{V}_{av}}$), as defined in equations (\ref{nd1})-(\ref{nd4}). For an overall picture, analysis of the full computational domain (seeFigure \ref{fig5}) reveals that $\mathrm{{W}^{V}_{av}}$, $\mathrm{|\Gamma|^{V}_{av}}$, and $\mathrm{{(J/B)}^{V}_{av}}$ decrease with time, indicating magnetic relaxation. For a more detailed analysis, we select three sub-volumes of interest, namely $\rm{S_{1}}$, $\rm{S_{2}}$, and $\rm{S_{3}}$ (see Figure \ref{t1}), where $\rm{S_{1}}$ is centered on the X-point of the HFT, $\rm{S_{2}}$ focuses on the HFT configuration, and $\rm{S_{3}}$ covers the full spatial extent of observed flaring region. To investigate the dynamics, we divide the simulation time into five phases, labeled by $\mathrm{P_{1}^{(i)}}$, $\mathrm{P_{2}^{(i)}}$, and $\mathrm{P_{3}^{(i)}}$ (where $\rm{i=1,2,...,5}$) for sub-volumes $\rm{S_{1}}$, $\rm{S_{2}}$, and $\rm{S_{3}}$, respectively.
Broadly, we find that in all sub-volumes the final values of $\mathrm{|\Gamma|^{V}_{av}}$ and $\mathrm{{(J/B)}^{V}_{av}}$ are smaller than their initial values, indicating a reduction in both twist and field gradient---consistent with the scenario of relaxation. Further, common to all sub-volumes, a sudden drop in $\mathrm{{J}^{V}_{av}}$ during the initial phase (e.g. $\mathrm{P_{1}}^{(1)}$ for $\rm{S_{1}}$) is governed prominently by layers adjacent to bottom boundary, indicating a possible boundary condition effect. 

Unlike the magnetic energy averaged over the whole computational grid, its grid averages over the sub-volumes do not decay monotonically. Toward understanding variations of magnetic energy in different sub-volumes, properties of $|{\bf{D}}|$, Poynting flux, and magnetic twist in each sub-volume are explored. Importantly, large values of $|{\bf{D}}|$ are found to be localized at $\rm{S_{1}}$, particularly coinciding with the location of the X-point of HFT. This is harmonious with the spirit of ILES. 
In $\rm{S_{1}}$, apart from the phase $\mathrm{P_{1}}^{(2)}$ and briefly in $\mathrm{P_{1}}^{(3)}$, the magnetic energy evolution is in conformity with physical expectations. The disagreement could be because of a failure of idealized Ohm's law as the induction equation in its ideal limit is not satisfied. Similar analyses have been carried out to explore energy variations in $\rm{S_{2}}$ and $\rm{S_{3}}$ also. In $\rm{S_{2}}$, the energy influx along with twist overwhelms dissipation, thus resulting in the observed energy increase from  $\mathrm{P_{2}}^{(3)}$ to  $\mathrm{P_{2}}^{(5)}$. The decrease in magnetic energy in $\rm{S_{3}}$ is found to be due to non-ideal effects primarily localized in $\rm{S_{1}}$.
Toward an estimation of the extent of relaxation, the angle between the current density and magnetic field ($\theta$) at every voxel is calculated. When integrated over the whole computational domain, the grid averaged angle drops by $4.3^{\circ}$. For the sub-volumes, it is found that the changes in $\theta$ distribution over the course of simulation are not very clear for $\rm{S_{1}}$ due to its small size. In $\rm{S_{2}}$ and $\rm{S_{3}}$, the peak of $\theta$ distribution becomes smaller, as realized from the decrease in fraction of voxels having $\rm{\theta}\geq 60^{\circ}$. The decrease in higher values of $\theta$ indicates increase in alignment between current density and magnetic field. 

In tandem, the above results indicate an ongoing magnetic relaxation, though it's extent remains to be explored. The angular distribution between current density and magnetic field suggests that though there is magnetic relaxation, but not enough to reach a force-free state. The terminal state of the simulation remains in non-equilibrium, suggesting the possibility for further relaxation. An auxiliary simulation with twice the computational time shows further alignment  of electric current density with magnetic field, but at a slower rate. This is expected as the corresponding time span overlaps with the observed post-flare phase where reconnection plays a secondary role. Overall, the simulation suggests the extent of solar fare induced magnetic relaxation depends on the flare energetics and its duration.  
To further contemplate, magnetic reconnections are localized in the flaring regions. Consequently, a flaring region can exchange magnetic helicity with its surroundings. Under such circumstances, invariance of helicity is non-trivial and a complete field alignment of electric current density may not be achieved. An explicit calculation of magnetic helicity and understanding its evolution is necessary to focus on this idea---which we leave as a future exercise.


%
%
\bibliography{main}

\IfFileExists{\jobname.bbl}{} {\typeout{}
\typeout{****************************************************}
\typeout{****************************************************}
\typeout{** Please run "bibtex \jobname" to obtain} \typeout{**
the bibliography and then re-run LaTeX} \typeout{** twice to fix
the references !}
\typeout{****************************************************}
\typeout{****************************************************}
\typeout{}}

\end{article}
\end{document}